\documentclass[fleqn,usenatbib]{mnras}

\usepackage{newtxtext,newtxmath}
\usepackage[T1]{fontenc}

\DeclareRobustCommand{\VAN}[3]{#2}
\let\VANthebibliography\thebibliography
\def\thebibliography{\DeclareRobustCommand{\VAN}[3]{##3}\VANthebibliography}

\usepackage{graphicx}	% Including figure files
\usepackage{amsmath}	% Advanced maths commands
\usepackage{xcolor}
\usepackage{subcaption}
\usepackage{orcidlink}
\newcommand{\km}{\rm\thinspace km}

\newcommand{\erg}{\rm\thinspace erg}
\newcommand{\s}{\rm\thinspace s}
\newcommand{\Mpc}{\rm\thinspace Mpc}

\newcommand{\kmps}{\hbox{$\km\s^{-1}$}}
\newcommand{\ergps}{\hbox{$\erg\s^{-1}\,$}}

\newcommand{\chisq}{\hbox{$\chi^2$}}

\newcommand{\kmpspMpc}{\hbox{$\kmps\Mpc^{-1}$}}
\newcommand{\civ}{\ion{C}{iv}}
\newcommand{\mgii}{\ion{Mg}{ii}}

\newcommand{\qsosed}{{\sc qsosed}}

\title[\ion{C}{IV} and \ion{He}{II} emission in quasars]{Testing AGN outflow and accretion models with \ion{C}{IV} and \ion{He}{II} emission line demographics in $z\approx2$ quasars}

\author[M. J. Temple et al.]{%
Matthew J. Temple$^{\orcidlink{0000-0001-8433-550X}}$,$^{1}$\thanks{E-mail: Matthew.Temple@mail.udp.cl}
James H. Matthews$^{\orcidlink{0000-0002-3493-7737}}$,$^{2,3}$
Paul C. Hewett$^{\orcidlink{0000-0002-6528-1937}}$,$^{3}$
Amy L. Rankine$^{\orcidlink{0000-0002-2091-1966}}$,$^{4}$
\newauthor 
Gordon T. Richards$^{\orcidlink{0000-0002-1061-1804}}$,$^{5}$
Manda Banerji$^{\orcidlink{0000-0002-0639-5141}}$,$^{6}$
Gary J. Ferland$^{\orcidlink{0000-0003-4503-6333}}$,$^{7}$
Christian Knigge$^{\orcidlink{0000-0002-1116-2553}}$$^{6}$
and
Matthew Stepney$^{\orcidlink{0000-0002-7711-0537}}$$^{6}$
\\
$^{1}$Instituto de Estudios Astrof\'{\i}sicos, Universidad Diego Portales, Av. Ej\'ercito Libertador 441, Santiago 8370191, Chile\\
$^{2}$Department of Physics, Astrophysics, University of Oxford, Denys Wilkinson Building, Keble Road, Oxford OX1 3RH, UK\\
$^{3}$Institute of Astronomy, University of Cambridge, Madingley Road, Cambridge CB3 0HA, UK\\
$^{4}$Institute for Astronomy, University of Edinburgh, Royal Observatory, Blackford Hill, Edinburgh EH9 3HJ, UK\\
$^{5}$Department of Physics, Drexel University, 32 S. 32nd Street, Philadelphia, PA 19104, USA\\
$^{6}$School of Physics \& Astronomy, University of Southampton, Southampton SO17 1BJ, UK\\
$^{7}$Department of Physics and Astronomy, The University of Kentucky, Lexington, KY 40506, USA
}

\date{Submitted to MNRAS 2023 April 24}

\pubyear{2023}

\begin{document}
\label{firstpage}
\pagerange{\pageref{firstpage}--\pageref{lastpage}}
\maketitle

\begin{abstract}
Using $\approx$190,000 spectra from the seventeenth data release of the Sloan Digital Sky Survey, we investigate the ultraviolet emission line properties in $z\approx2$ quasars.
Specifically, we quantify how the shape of \ion{C}{IV}\,$\lambda$1549 and the equivalent width (EW) of  \ion{He}{II}\,$\lambda$1640 depend on the black hole mass and Eddington ratio inferred from \ion{Mg}{II}\,$\lambda$2800.
 Above $L/L_\textrm{Edd}\gtrsim0.2$, there is a strong mass dependence in both \ion{C}{IV} blueshift and \ion{He}{II} EW.  Large \ion{C}{IV} blueshifts are observed only in regions with both high mass and high accretion rate.
Including X-ray measurements for a subsample of 5,000 objects, we interpret our observations in the context of AGN accretion and outflow mechanisms.
The observed trends in \ion{He}{II} and 2\,keV strength are broadly consistent with theoretical \qsosed\ models of AGN spectral energy distributions (SEDs) for low spin black holes, where the ionizing SED depends on the accretion disc temperature and the strength of the soft excess. High spin models are not consistent with observations, suggesting SDSS quasars at $z\approx2$ may in general have low spins.
We find a dramatic switch in behaviour at $L/L_\textrm{Edd}\lesssim0.1$: the ultraviolet emission properties show much weaker trends, and no longer agree with \qsosed\ predictions, hinting at changes in the structure of the broad line region.
Overall the observed emission line trends are generally consistent with predictions for radiation line driving where quasar outflows are governed by the SED, which itself results from the accretion flow and hence depends on both the SMBH mass and accretion rate.
\end{abstract}
% 246 words


%%%%%%%%%%%%%%%%%%%%%%%%%%%%%%%%%%%%%%%%%%%%%%%%%%

%%%%%%%%%%%%%%%%% BODY OF PAPER %%%%%%%%%%%%%%%%%%

\section{Introduction}

\subsection{Observational context: spectroscopic properties of quasars}

The spectroscopic properties of  type-1 quasars have long been appreciated  for their potential to provide insight into the physical processes  responsible for luminous active galactic nuclei \citep[AGN;][]{1978ApJ...226....1B, 1979RvMP...51..715D, 1981ApJ...250..478K, 1988ApJ...324..714K, 2000ApJ...545...63E}.
These processes include the excitation of various line- and continuum-emitting regions, and  mechanisms for launching outflows which might `feed back' energy to their host galaxies.
Such processes are  ultimately powered by accretion onto supermassive black holes (SMBHs; \citealt{1969Natur.223..690L}), and thus depend primarily on the mass of the SMBH and the accretion rate, with potential second-order drivers including the spin of the SMBH and the metal content of the accreting material.

The search for insight has gained much from identifying and exploring the properties which are observed to vary the most. 
Such diversity in the observed quantities must ultimately be driven by some of the physics which we would like to use to better constrain both the growth of SMBHs and their effect on the galactic ecosystems in which they reside.
The most famous result of these investigations is arguably the identification of the so-called `eigenvector 1' (EV1), which accounts for the largest amount of correlated variance in the optical spectra of low-redshift ($z<1$) type-1 AGN spectra.
Most authors now agree that the EV1 is driven by the mass-normalised accretion rate (the Eddington ratio), possibly convolved with some orientation effect \citep{1992ApJS...80..109B, 1999ApJ...515L..53W, 2000ApJ...536L...5S, 2014Natur.513..210S, 2015ApJ...804L..15S, 2015FrASS...2....6S, 2020MNRAS.492.3580W}.

Similarly, the ultraviolet emission features in quasar spectra also show a rich phenomenology \citep{2002MNRAS.337..275C, 2016ApJ...833..199J, 2022AJ....163..110B}.
Early work by \citet{1977ApJ...214..679B} showed that the equivalent widths (EWs) of various ultraviolet lines, most notably \ion{C}{IV}\,$\lambda$1549, were anti-correlated with the ultraviolet continuum luminosity.
\citet{2003ApJ...586...52S} showed that this `Baldwin effect' was independent of EV1 in 22 quasars with $z<0.4$, implying different physical drivers for these correlations.
Early observations also demonstrated that the centroid of the \ion{C}{IV} emission line is commonly shifted to the blue \citep{1982ApJ...263...79G, 1984MNRAS.207...73W, 2002AJ....124....1R}. 
Within a sample of 87 Palomar-Green quasars, \citet{2004MNRAS.350L..31B, 2005MNRAS.356.1029B} found that large \civ\ blueshifts were only seen in objects with high Eddington ratios, although not all quasars with high Eddington ratios had large \civ\ blueshifts.
The EV1 formalism was extended by \citet{2004ApJ...617..171B} and \citet{2007ApJ...666..757S} to include the velocity shift of \ion{C}{IV}, again finding that large \ion{C}{IV} blueshifts are seen only in so-called `Population A' quasars with high Eddington ratios.

With the start of the Sloan Digital Sky Survey \citep[SDSS;][]{2000AJ....120.1579Y}, large samples of rest-frame ultraviolet quasar spectra became available.
Equally as important were methods to accurately characterize the systemic redshift of each quasar \citep{2010MNRAS.405.2302H}, which is necessary to infer the velocity shift of any emission features.
A notable work by \citet{2011AJ....141..167R} summarized the state of the field a decade ago at the time of the seventh data release \citep[DR7;][]{2010AJ....139.2360S} from SDSS. Using $\approx$35\,000 quasar spectra, \citet{2011AJ....141..167R} confirmed the Baldwin effect and showed that the EW of \ion{C}{IV} line also anti-correlates with the  magnitude of the \ion{C}{IV}  blueshift: quasars with higher luminosities show, on average, weaker \ion{C}{IV} emission which is more strongly blueshifted. 
\civ\ blueshifts could be a signature of emission from ionized gas being driven away from the accretion disc along the line-of-sight to the observer \citep{2004ApJ...611..107L}, in which case the results of \citet{2011AJ....141..167R} can be interpreted as brighter objects showing stronger emission from outflowing gas and weaker emission from the virialized broad line region (BLR). We discuss this interpretation further in Section~\ref{sec:discuss_outflows}, but do not assume anything about the origin of \civ\ line shifts when presenting our observational results in Section~\ref{sec:results:obs}.

\citet{2011AJ....141..167R} also demonstrated that the \ion{C}{IV} properties are strongly correlated with the EW of the nearby \ion{He}{II}\,$\lambda$1640 emission line. 
More recent work by \citet{2020MNRAS.492.4553R} has shown that the correlations between the EW of \ion{He}{II} and both the EW and blueshift of \ion{C}{IV} are also present in quasars with broad absorption features. We now know that the \ion{C}{IV} and \ion{He}{II} properties are strongly correlated with the properties of other ultraviolet emission features such as Ly\,$\alpha$, \ion{N}{V}, \ion{Si}{IV} and \ion{O}{IV}] \citep{2021MNRAS.505.3247T},
 \ion{Fe}{III}, \ion{Al}{III}, \ion{Si}{III}] and \ion{C}{III}] \citep{2020MNRAS.496.2565T},
 as well as the optical [\ion{O}{III}] emission  \citep{2018A&A...617A..81V, 2019MNRAS.486.5335C, 2020A&A...644A.175V}, the strength of near infrared emission from dust at the sublimation temperature \citep{2021MNRAS.501.3061T}, 
 the strength of the far infrared emission \citep{2017MNRAS.470.2314M},
 the radio detection fraction \citep{rankine2021}
 and the strength of the 2\,keV X-ray continuum 
 \citep{2011AJ....142..130K, 2020A&A...635L...5Z, 2020MNRAS.492..719T, 2021MNRAS.504.5556T, 2021A&A...653A.158L, 2022ApJ...931...41M, 2022ApJ...931..154R}.
 Tentative links have also been found between the \civ\ blueshift and the amount of continuum reddening ascribed to nuclear dust \citep{2021A&A...649A.102C, 2022MNRAS.513.1254F}.

The existence of such correlations - between parameters which trace emission at different wavelengths and from different physical regions  - suggests that they are driven (either directly or indirectly) by changes in some of the fundamental physical parameters which govern the properties of a SMBH and its surrounding regions, such as the SMBH mass, spin, and accretion rate.
 The space spanned by \civ\ blueshift and \civ\ EW therefore appears to be just as important as EV1 in understanding the physics of luminous AGN.
However, while the location of a given quasar spectrum on either EV1 or the \ion{C}{IV} blueshift--EW space must ultimately be a function of the fundamental SMBH parameters, there is no guarantee that such a function is linear, or even injective (i.e.\,one-to-one with a well-defined inverse). For example, we cannot rule out the possibility that two objects with different $M_\textrm{BH}$ and accretion rate have similar (or indeed identical) \ion{C}{IV} emission.
%Possible analogy with stars mapping onto HR diagram - SMBH accretion process are more complicated than isolated ZAMS stars and so, for all we would like to use EV1 as a HR-like diagram, there isn't a good reason to expect it to be as simple.

\subsection{Theoretical context: AGN outflows and SEDs}
\label{sec:intro:outflows}

Mass outflows from AGN can be launched by thermal pressure, magnetic forces, or radiation \citep[][]{2021NatAs...5...13L}.
Thermal winds can only be launched at large radii [$R \gtrsim 10^5 R_\textrm{g} \approx 5 \times (M_\textrm{BH}/10^9 M_\odot)$\,parsec] with terminal velocities of order 100--1000\,\kmps\ 
\citep{1983ApJ...271...70B, 1996ApJ...461..767W, 2019MNRAS.489.1152M}.
Faster outflows with speeds $>2000$\,\kmps, as commonly seen in broad high-ionization  ultraviolet absorption features,
are most likely launched on smaller (sub-parsec) scales.
Magnetically driven winds \citep{1982MNRAS.199..883B, 1992ApJ...385..460E, 1994ApJ...434..446K, 2010ApJ...715..636F, 2021ApJ...914...31Y} may be important in this context, but we currently lack predictive models for how such winds would translate into observable quantities \citep[although see e.g.][]{bottorff2000,chajet2013}.
On the other hand, radiation line driving \citep{1975ApJ...195..157C, 1995ApJ...451..498M, 1995ApJ...454L.105M, 2000ApJ...545...63E, 2000ApJ...543..686P, 2004ApJ...616..688P, 2007ApJ...661..693P, 2010A&A...516A..89R, 2017ApJ...847...56E, 2017MNRAS.465.2873N, 2020MNRAS.494.3616N, 2022MNRAS.513.1141Z} is intrinsically linked to the spectral energy distribution (SED) of the continuum which is responsible for both ionizing the transitions and then accelerating the flow by providing the source of radiation pressure.
By considering how the SED changes with SMBH mass and accretion rate, authors such as \citet{2019A&A...630A..94G} have developed unifying frameworks which make testable predictions for luminous AGN based on the physics of radiation line driven winds.

The ionizing continuum SED depends on the structure of the accretion flow, which in turn is set by the SMBH mass $M_\textrm{BH}$, the mass-normalised accretion rate $\dot{m} = \dot{M}_\textrm{BH}/\dot{M}_\textrm{Edd}$ and the SMBH spin $a_*$.
Empirically, the optical--to--X-ray SEDs of AGN are seen to contain at least three  distinct components
\citep[][]{1994ApJS...95....1E, 2006ApJ...637..157C, 2007ApJS..173....1L, 2012MNRAS.420.1848D, 2012MNRAS.420.1825J}. %, which are most likely emitted from physically distinct locations.
First, any optically thick accretion disc will emit thermally, with larger radii being cooler, giving rise to a multi-temperature blackbody which is expected to peak in the near-ultraviolet ($M_\textrm{BH} > 10^8 M_\odot$), far-ultraviolet or soft X-rays ($M_\textrm{BH} < 10^8 M_\odot$).
This peak would be expected to depend on $M_\textrm{BH}$ if larger SMBHs have accretion flows which truncate at lower temperatures \citep[eq.~5.3.1 of][]{1973blho.conf..343N}.
However, this part of the SED is instead commonly observed to peak around 1100\,\AA\ \citep[][]{2005ApJ...619...41S, 2013ApJS..206....4K, 2014ApJ...794...75S, 2020MNRAS.493.2745V}, albeit with lower luminosity AGN showing harder continuum emission at $\lambda < 1100$\,\AA\ \citep{2002ApJ...579..500T, 2004ApJ...615..135S}.
The uniformity of this peak wavelength across a wide range of $M_\textrm{BH}$ and $\dot{m}$ has been suggested to result from opacity effects \citep{1987ApJ...321..305C}
or from line-driven winds which remove mass from the inner accretion disc \citep{2012MNRAS.426..656S, 2014MNRAS.438.3024L}.
Second, a hot Comptonised `corona' emits a non-thermal power law which dominates the X-ray continuum above 1\,keV \citep{1991ApJ...380L..51H, 1994ApJ...434..570T}.
Finally, a `soft excess' is seen in the X-rays below $\approx$1\,keV, which is usually attributed to an intermediate warm Comptonising component \citep{2018A&A...611A..59P, 2020A&A...634A..85P}. 
This soft excess may be a significant contributor to the ionizing  SED in the $\approx$100--1000\,\AA\ ($\approx$10--100\,eV) range, where many of the ultraviolet transitions are excited, but where direct observations of the continuum are not possible due to the high opacity of neutral hydrogen along the line-of-sight.

For a line-driven disc-wind to emerge, the system needs strong ultraviolet emission to produce sufficient radiation pressure, but also a soft enough SED to avoid over-ionizing the gas \citep{1995ApJ...451..498M, 2014ApJ...789...19H}.
For each relevant line, the flux at the line energy combined with the line opacity  determines the line-driving boost beyond radiation pressure from Thomson scattering. 
The line opacity depends on the ionization state,  which is primarily sensitive to the flux at the ionization edges 
(48 and 64\,eV for the production and destruction of \ion{C}{IV}).
Line driving results when this effect is summed across many lines, each with their own energies, leading to a complex interplay between the flux of the SED underneath all the relevant lines in the ultraviolet  and the flux of the SED beyond all the relevant ionization edges.
\citet{2019A&A...630A..94G} suggest that both $L/L_\textrm{Edd} \gtrsim 0.25$ and $M_\textrm{BH} \gtrsim 10^8 M_\odot$ are required to satisfy these criteria and hence to power a strong outflow through radiation line driving. 
 \citet{2019A&A...630A..94G} also expect the $M_\textrm{BH}$ dependence of the observed outflow properties to be different above and below an $\dot{m}$ of around 0.25, 
where they expect the cold, optically thick accretion disc to extend down towards the innermost stable circular orbit (ISCO) and replace the hot, optically thin, inner accretion flow which is present at lower accretion rates.
In other words, they require $L/L_\textrm{Edd} \gtrsim 0.25$ to ensure emission from thermal disc emission dominates over that from the hot corona, to accelerate a strong line-driven wind without over-ionizing the gas.
With the quantity and quality of spectroscopic data which are now available from large surveys, these predictions from the \citet{2019A&A...630A..94G} framework can be tested empirically.

\subsection{Observational probes of quasar SEDs}

From an observational viewpoint, it is relatively easy to constrain the SED of an unobscured type-1 AGN in the rest-frame infrared, optical and X-ray wavebands, 
as photometric measurements can place direct constraints on the emission.
For example, the strength of the rest-frame 2\,keV X-ray continuum relative to the near-ultraviolet continuum has been shown to anti-correlate with the ultraviolet continuum luminosity in the so-called  $\alpha_\textrm{ox}$--$L_\textrm{2500\,\AA}$ relation 
\citep{1982ApJ...262L..17A, 1986ApJ...305...83A, 2006AJ....131.2826S, 2007ApJ...665.1004J, 2016ApJ...819..154L, 2021MNRAS.504.5556T},
and the fractional contribution of the 2-10\,keV emission to the total bolometric power of the AGN is known to vary as a function of the accretion rate $\dot{m}$ \citep{2007MNRAS.381.1235V, 2009MNRAS.392.1124V}.
However, while this X-ray waveband can make an important contribution to the total emitted energy, it contributes a negligible number of ionizing photons to the photoionization budget of the BLR gas (see Appendix~\ref{appendixa}).
The number of ionizing photons is instead dominated by photons at the ionization edges themselves, which is of the order of 10-100\,eV for the ultraviolet BLR 
(e.g. production and destruction edges of 7.6 and 15\,eV for \ion{Mg}{II}, 48 and 64\,eV for \civ, and 24 and 54\,eV for \ion{He}{II}; see also fig.~13 of \citealt{2011AJ....141..167R}).
This extreme ultraviolet (EUV) part of the SED is not directly observable due to intervening absorption along the line-of-sight, but plays a crucial role in the physics of the BLR.
To add to the complexity, the relative contribution of the  warm Comptonising soft excess to the total EUV emission is likely to be varying as a function of $M_\textrm{BH}$ and $\dot{m}$, meaning that the observable 2\,keV continuum may not be a reliable proxy for the strength of the EUV SED at the ionization edges.

\ion{C}{IV} is a resonant doublet transition with a complicated ionic structure, so the strength of \ion{C}{IV} emission is not necessarily a good tracer of the ionizing SED.
However, it is instead possible to probe the EUV continuum using the \ion{He}{II}\,$\lambda$1640 recombination line, which arises from a simple hydrogenic (i.e.\,single electron) system.
Under the assumptions that the \ion{He}{II} emitting region is in equilibrium and that the \ion{He}{II} continuum is optically thick, the total rate of \ion{He}{II}-ionizing photons must balance the total number of recombinations such that each \ion{He}{II}\,$\lambda$1640 line photon can be associated with an ionizing continuum photon at or above 54\,eV.
This method was first used by \citet{zanstra1929} to infer stellar temperatures using the strength of Hydrogen recombination lines \citep[section 5.10 of][]{2006agna.book.....O}.
Following previous works \citep{1987ApJ...323..456M, 2013MNRAS.432.1525B, 2020MNRAS.494.5917F, 2021MNRAS.504.5556T}, we will use the strength of \ion{He}{II} as a proxy for the strength of the `unseen' EUV continuum which is ionizing the BLR.

\subsection{This work}
\label{sec:work}
The first goal of this paper is to provide an up-to-date summary of our knowledge of the ultraviolet spectral properties of type-1 quasars,  using  the final data release (DR17) from the fourth iteration of SDSS. This sample contains an order of magnitude more quasars than the SDSS DR7 sample used by \citet{2011AJ....141..167R}. The large sample size allows us to consider the emission properties as a function of both mass and Eddington ratio simultaneously, and thus provide a test of 
 {model quasar SEDs from \citet{2018MNRAS.480.1247K} and of current theories of radiation line-driven AGN winds \citep{2019A&A...630A..94G}}, which is our second goal.

To best compare with theory and simulations, we present observed quantities such as the \ion{C}{IV} blueshift, \ion{He}{II} EW, and $\alpha_\textrm{ox}$ as a function of three physical parameters: the ultraviolet continuum luminosity, the SMBH mass estimated from the \ion{Mg}{II}\,$\lambda$2800 emission line, and the inferred Eddington ratio.
This relatively simple exercise has long been used to gain insight into the physics of AGN \citep[][]{1980SvA....24..389D}, 
but is subtly different from purely empirical approaches which  observe trends in emission line properties (e.g. EV1 or the \ion{C}{IV} blueshift-EW plane) and then try to infer which underlying physical parameters are driving those trends.
By contrast, theoretical models make predictions for the SED and outflow properties as a function of the SMBH mass and accretion rate.
In this work we confront such predictions directly with observations, 
showing that the ultraviolet emission lines display different behaviour above a threshold of $L_\textrm{bol}/L_\textrm{Edd}\approx 0.2$, consistent with predictions for radiation line-driven winds, and finding good qualitative agreement between 
the \citet{2018MNRAS.480.1247K} SED models and observed continuum tracers in regions of parameter space where their models were not calibrated.

The structure of this paper is as follows.
In Section~\ref{sec:data} we present the observational data, while in Section~\ref{sec:models} we describe the SED models to which we compare.
We present our key results in Section~\ref{sec:results} and discuss their implications and limitations in Section~\ref{sec:discuss}.
Throughout this work, wavelengths are given in vacuum in units of \AA ngstr\"oms, and we assume a flat $\Lambda$CDM cosmology with $\Omega_m=0.27$, $\Omega_\Lambda=0.73$ and $H_0=71$\kmpspMpc. 
Energies, frequencies and wavelengths are given in the rest-frame unless stated otherwise.
%We use the notation $L_\textrm{2\,keV}$, $L_\textrm{2500}$ and $L_\textrm{3000}$ to refer to the monochromatic continuum luminosities $\nu L_\nu \;( = \lambda L_\lambda)$ in units of \ergps at 2\,keV, 2500\,\AA\ and 3000\,\AA\ respectively.

\section{Observational data}
\label{sec:data}

\begin{figure*}
 \begin{subfigure}{0.75\columnwidth}
	\includegraphics[width=\columnwidth]{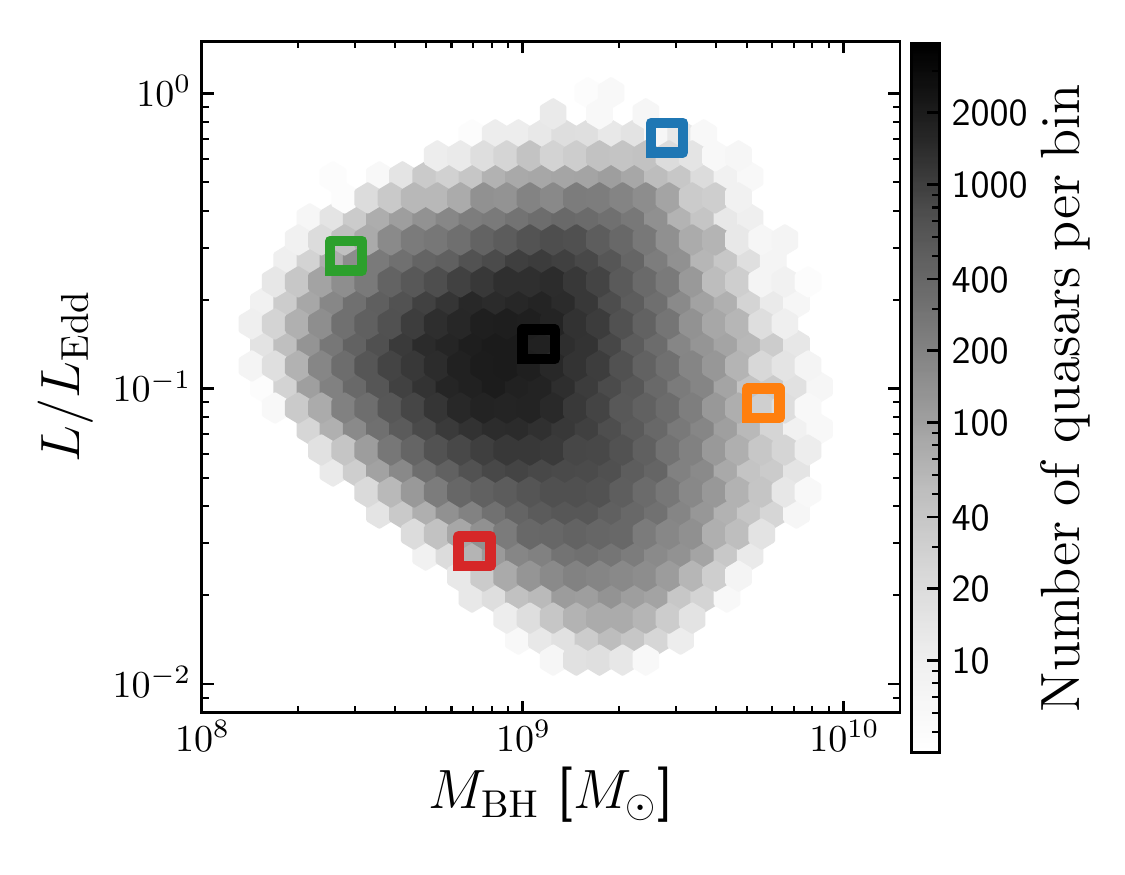}
	\includegraphics[width=\columnwidth]{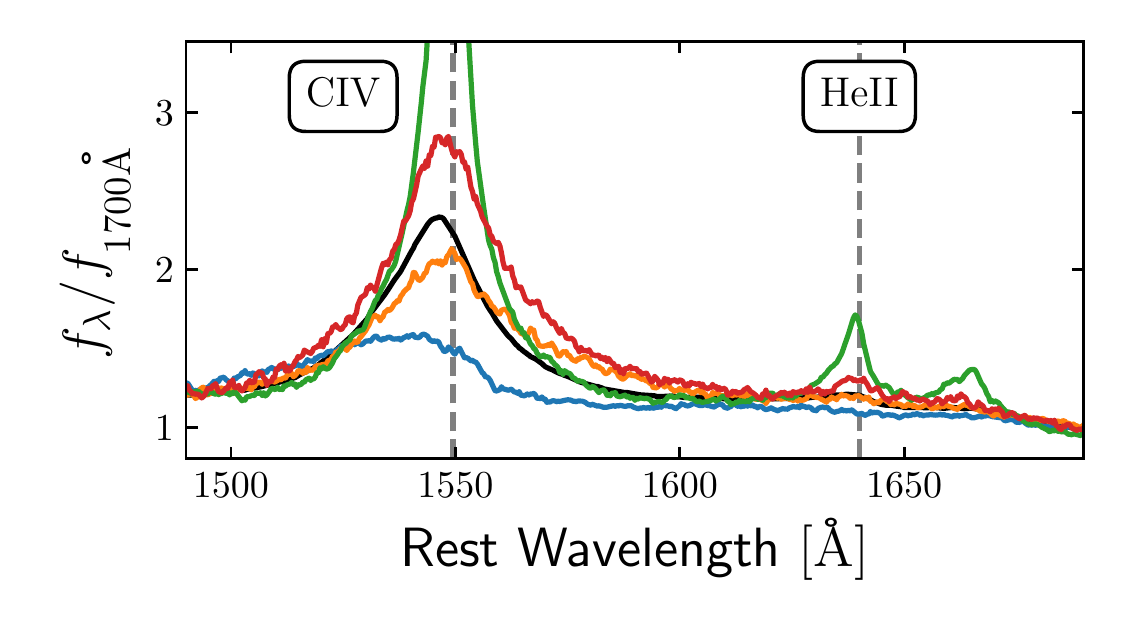}
 \end{subfigure}
 \begin{subfigure}{1.3\columnwidth}
	\includegraphics[width=\columnwidth]{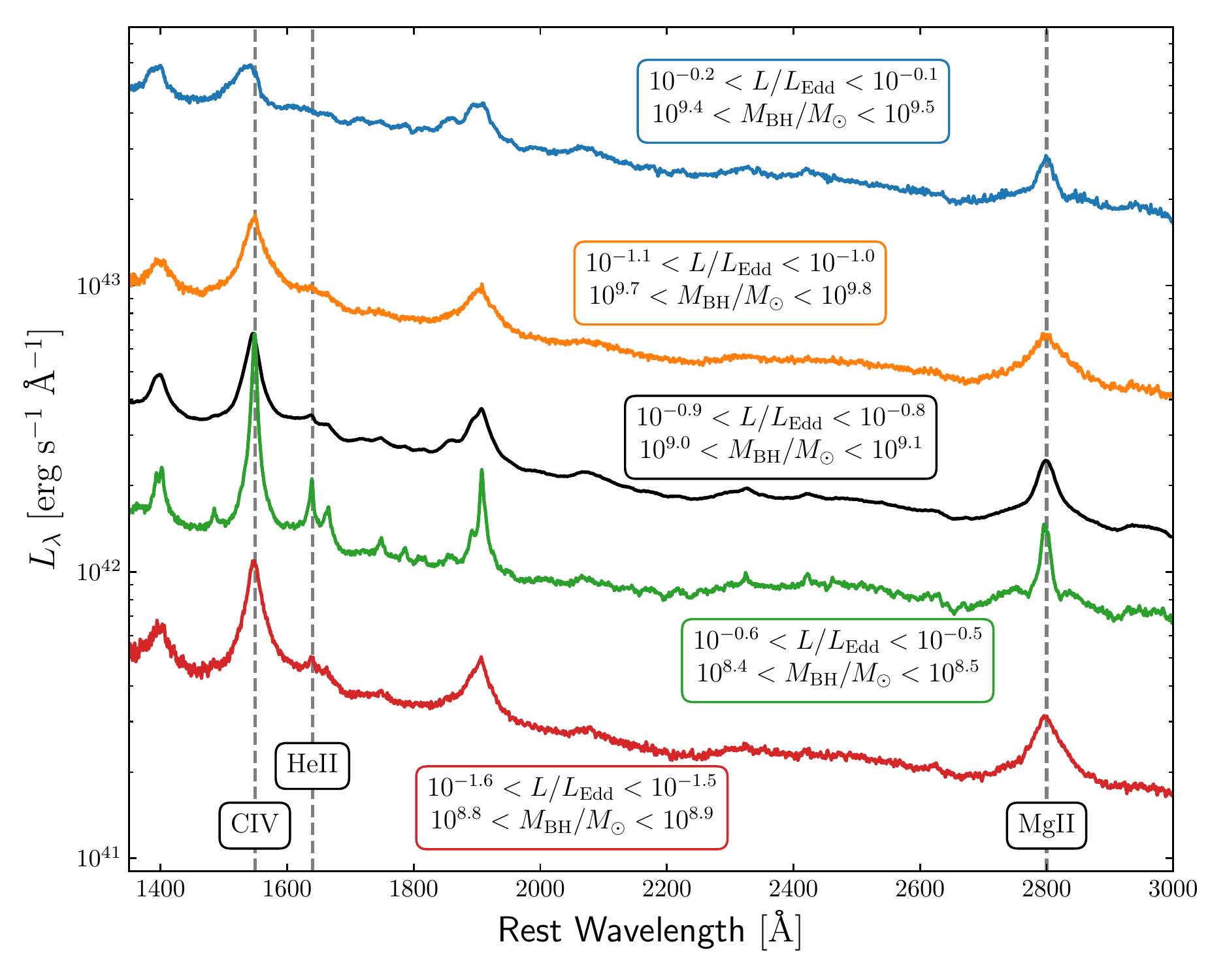}
 \end{subfigure}
    \caption{%This is an example figure. Captions appear below each figure.
	%Give enough detail for the reader to understand what they're looking at,
	%but leave detailed discussion to the main body of the text.
	\textsl{Top left panel:} the distribution of our sample of 186\,303 quasars with redshifts $1.5< z < 2.65$ in the $M_\textrm{BH}$--$L/L_\textrm{Edd}$ plane. Throughout this work, we only consider hexagonal bins where there are five or more quasars per bin.
    By construction, the FWHM of \ion{Mg}{II}\,$\lambda$2800 increases from top-left to bottom-right of this parameter space, while the 3000\,\AA\ continuum luminosity increases from bottom-left to top-right.
    \textsl{Right panel:} composite spectra taken from the different regions of the $M_\textrm{BH}$--$L/L_\textrm{Edd}$ plane indicated in the top left panel. 
     {In black is shown a composite from the densely populated region in the centre of the $M_\textrm{BH}$--$L/L_\textrm{Edd}$ plane. The coloured composites were chosen to illustrate the full diversity of emission line properties which can be seen with changing SMBH mass and Eddington ratio.} 
    \textsl{Bottom left panel:} comparing the composite spectra in the region around \civ\,$\lambda$1549 and \ion{He}{II}\,$\lambda$1640. Here the spectra have been normalised at 1700\,\AA\ and plotted on a linear y-axis.
    The EW of \ion{He}{II} can be seen to correlate with the profile of \civ: the high-mass, high-Eddington composite in blue displays weak lines and blueshifted \civ\ while the low-mass, high-Eddington composite in green shows much stronger line emission with no blue excess in \civ, consistent with fig. 11 of \citet{2011AJ....141..167R} and fig. A2 of \citet{2020MNRAS.492.4553R}. The difference here is that, instead of being constructed from \civ\ or \ion{C}{III}] emission properties, objects were included based on the FWHM of \ion{Mg}{II} and $L_{3000}$ to represent regions of the $M_\textrm{BH}$--$L/L_\textrm{Edd}$ plane, and also that the larger  sample from SDSS DR17 includes fainter objects such as those contributing to the composite in red.
    Composite spectra spanning the full range of the $M_\textrm{BH}$--$L/L_\textrm{Edd}$ space are available as supplemental online-only material with the journal.
	}
    \label{fig:sample}
\end{figure*}

\subsection{Rest-frame ultraviolet spectra}

The first aim of this paper is to quantify the behaviour of \ion{He}{II}\,$\lambda1640$ and \ion{C}{IV}\,$\lambda 1549$ as a function of SMBH mass $M_\textrm{BH}$ and Eddington ratio $L/L_\textrm{Edd}$.
The $M_\textrm{BH}$ inferred from single-epoch measurements of \ion{C}{IV} is known to be biased as a function of the emission line properties \citep{2005MNRAS.356.1029B, 2008ApJ...680..169S, 2016MNRAS.461..647C, 2017MNRAS.465.2120C, 2018MNRAS.478.1929M},  so we will instead use the velocity width of the \ion{Mg}{II}\,$\lambda 2800$ line to infer $M_\textrm{BH}$.
We construct a sample of quasars from the SDSS with coverage of rest-frame wavelengths 1450-3000\,\AA\ to include \ion{C}{IV}, \ion{He}{II} and \ion{Mg}{II} (Fig.~\ref{fig:sample}). 

The original selection of the SDSS DR17 quasar sample was described by \citet{2020ApJS..250....8L} and \citet{2022ApJS..259...35A}. 
We post-process each spectrum using a sky subtraction routine conceptually similar to  that described by \citet{2005MNRAS.358.1083W}\footnote{Measurements of spectrum properties derived from observed-frame wavelengths $>$6700\,\AA \ improve somewhat but none of the results, or conclusions, of this paper change if the original DR17 reductions of the spectra are used instead.}.
Systemic redshifts are calculated as described in section 3 of \citet{2020MNRAS.492.4553R}. Our redshift estimation routine uses the rest-frame 1600-3000\,\AA\ region, deliberately excluding the \ion{C}{IV} emission line, which is a key difference compared to the approach employed in the SDSS quasar catalogues.
The improved redshifts and sky-subtracted spectra will be described in a forthcoming publication by P.\,C.\,Hewett.
To measure the emission line properties, we employ the spectral reconstructions from the Mean-Field Independent Component Analysis (ICA) carried out by \citet{2020MNRAS.492.4553R}, which we have successfully used in our previous investigations into quasar emission line physics \citep{2020MNRAS.496.2565T, 2021MNRAS.501.3061T, 2021MNRAS.505.3247T}. 
 {Ten spectral ICA components are used to reconstruct each spectrum, using an iterative routine to mask absorption features while fitting linear combinations of the components to the data.}
The ICA-reconstructions provide a significant improvement in the measurement of emission line properties, reducing the impact of the modest signal-to-noise ratio ($S/N$) in the original spectra and the effect of weak absorption lines (e.g. intervening or outflowing \ion{C}{IV}\,$\lambda\lambda$1548,1550 doublets).
We exclude objects with broad low-ionization absorption features, 
as such absorption features may affect the \ion{Mg}{II} emission line (in the case of LoBALs), or lead to a sub-optimal reconstruction of the \civ\ line (in the case of FeLoBALs).
A small fraction ($<1$ per cent) of the sample was excluded either because the ICA failed to converge, or because the resulting reconstruction had reduced $\chisq>2$, as described in section 4.3 of \citet{2020MNRAS.492.4553R}.
To include both \civ\ and \mgii\ in the observed spectrum, we limit our sample to redshifts $1.5 < z< 2.65$. Spectra from before the start of the BOSS survey (MJD 55000) were observed using the original SDSS spectrograph which had a more limited wavelength coverage; for these objects we require $1.6<z<2.2$ to ensure coverage of  \civ\ and \mgii.
Each quasar spectrum is required to possess a mean $S/N$ (per 69\kmps SDSS pixel) $\geq$3.0 over the rest-frame interval 1700-2200\,\AA.
These criteria leave a sample of 186\,303 quasars.

\civ\ and \ion{He}{II} emission properties are measured consistently with \citet{2011AJ....141..167R} and \citet{2020MNRAS.492.4553R}. 
To compute the EW of \ion{C}{IV} emission,  a power law continuum is defined using the median flux in the 1445-1465 and 1700-1705\,\AA\ wavelength windows.
This continuum is then subtracted from the spectrum to isolate the line flux in the 1500-1600\,\AA\ wavelength region.
The \ion{He}{II} EW is measured in the same way across the 1620-1650\,\AA\ wavelength region, using windows at 1610-1620 and 1700-1705\,\AA\ to define the continuum model.
The \ion{C}{IV} emission line `blueshift' is defined as the Doppler shift of the wavelength bisecting the continuum-subtracted line flux:
\begin{equation}
    \textrm{\ion{C}{IV} blueshift} \equiv c\times \left(\frac{\lambda_\textrm{rest}-\lambda_\textrm{median}}{\lambda_\textrm{rest}}\right)
\label{eq:blue}
\end{equation}
where $c$ is the speed of light, $\lambda_\textrm{median}$ is the rest-frame wavelength of the observed line centroid, and $\lambda_\textrm{rest}=1549.48$\,\AA \ is the mean rest-frame wavelength of the \ion{C}{IV}\,$\lambda\lambda$1548.19,1550.77 doublet.
% assuming equal contributions to the emission.

\subsection{X-ray data}
\label{sec:Xray_data}

In addition to the rest-frame ultraviolet emission features, we can use the rest-frame 2\,keV X-ray continuum emission to gain further insight into the SEDs of the quasars in our sample.
We therefore  cross-match our sample of 186\,303
objects to various X-ray catalogues from the literature, in order to build a large sample of rest-frame 2\,keV measurements.
4000 objects from our sample of $1.5<z<2.65$ objects with ultraviolet spectra are included in the recent study of quasar X-ray properties  by \citet{2022ApJ...931..154R},
including 
2691 with \textit{XMM-Newton} observations from \citet{2020A&A...642A.150L},
1291 with \textit{Chandra} observations from \citet{2020MNRAS.492..719T}, 
and 18 with \textit{XMM-Newton} observations from \citet{2020ApJS..250...32L}.
% 
% 215 sources from DR17. Rest of eFEDS sources from DR16Q (972-215 = 757).
972 sources in our sample have X-ray data from the \textit{eROSITA} Final Equatorial Depth Survey \citep[eFEDS;][]{2022A&A...661A...5L}: 757 objects from  the SDSS DR16 quasar catalogue \citep{2020ApJS..250....8L} and 
 215 sources with spectra released in SDSS DR17 \citep{2022ApJS..259...35A}.
We use data from the second ROSAT All-Sky Survey \citep[2RXS;][]{2016A&A...588A.103B} for objects included in the SDSS DR16 SPIDERS programme \citep{2017MNRAS.469.1065D, 2020A&A...636A..97C}. The flux limit for this survey is relatively bright  so we use the Bayesian  measurements described by  \citet{2019A&A...625A.123C} which account for the Eddington bias. A total of 36 objects from 2RXS are included in our sample.
Finally, we include 7, 9, and 7 objects with \textit{Chandra} observations from \citet{2021MNRAS.504.5556T}, \citet{2018MNRAS.480.5184N, 2022MNRAS.511.5251N} and \citet{2022ApJ...934...97F} respectively. These last three sub-samples were selected to have high $L_\textrm{UV}$, weak \ion{C}{IV} and strong \ion{C}{IV} respectively, but the number of quasars is small and our results would be unchanged if we were to exclude them.

The compilation results in a sample of 5031
quasars with measurements of both their ultraviolet (2500\,\AA) and X-ray (2\,keV) continuum fluxes. 
We use the rest-frame 2\,keV fluxes reported by  \citet{2018MNRAS.480.5184N, 2022MNRAS.511.5251N}, \citet{2019A&A...625A.123C},  \citet{2021MNRAS.504.5556T},  \citet{2022ApJ...934...97F},  \citet{2022A&A...661A...5L}, and \citet{2022ApJ...931..154R}, without any restriction on the spectral slope.
However, we have verified that the conclusions of this work would not change if we excluded objects which may be affected by absorption.
From these fluxes we compute luminosities assuming a consistent cosmology (Section~\ref{sec:work}) across all sub-samples.
We then infer $\alpha_\textrm{ox}$, 
the logarithm of the ratio of the rest-frame 2\,keV and 2500\,\AA\ monochromatic luminosities:
\begin{equation}
    \alpha_\textrm{ox}
    = \textrm{log}_{10}\Big(
    {\nu L_\nu}\Big)_\textrm{2\,keV} - \textrm{log}_{10}\Big({\nu L_\nu}\Big)_\textrm{2500\,\AA},
\end{equation}
as a measure of the relative strength of the X-ray emission in each source. 
Objects with smaller (i.e.\ more negative) $\alpha_\textrm{ox}$ have weaker 2\,keV X-ray emission relative to the ultraviolet continuum.

\subsection{Black hole masses and Eddington ratios}

We estimate SMBH masses using the single-epoch virial  estimator described by \citet{2009ApJ...699..800V}, using the full width at half maximum (FWHM) of the \ion{Mg}{II} line:
\begin{equation}
    M_\textrm{BH} = 10^{6.86}
    \left(\frac{\textrm{{\small FWHM}(\ion{Mg}{II})}}{1000\kmps}\right)^2
    \left(\frac{L_{3000}}{10^{44}\ergps}\right)^{0.5}M_\odot.
\end{equation}
where $L_{3000}$ is the rest-frame monochromatic continuum luminosity $\nu L_\nu$ at 3000\,\AA.
This $M_\textrm{BH}$ estimator assumes a relationship between the radius of the \ion{Mg}{II}-emitting region and the observed $L_{3000}$ which is independent of  the shape of the ionizing SED, or more generally, independent of any changes in the accretion disc structure which may arise with changing $M_\textrm{BH}$ or accretion rate. We discuss this assumption further in Section~\ref{sec:assumptions}.
We infer the FWHM of \ion{Mg}{II} from our ICA reconstructions, which provide a more robust model of the intrinsic \ion{Mg}{II} profile than a conventional Gaussian fit in low $S/N$ spectra.
Using a sub-sample with $S/N>10$,
we have verified that our \ion{Mg}{II} FWHM measurements are consistent with those obtained from fitting a single Gaussian to \ion{Mg}{II} together with an  iron template \citep{2001ApJS..134....1V} using the routine described by \citet{2011ApJS..194...45S}.
The key results of this paper would not change if we were to instead use such a Gaussian model for \ion{Mg}{II}, but there would be significantly more scatter in lower luminosity regions of parameter space where the spectral $S/N$ is poorer on average.
The error budget on our resulting $M_\textrm{BH}$ is dominated by the 0.55\,dex uncertainty on the single-epoch estimator as described by \citet{2009ApJ...699..800V}.

We infer $L_{3000}$ by fitting a quasar SED model \citep{2021MNRAS.508..737T} to \textit{griz} photometry. For sources in SDSS DR16 we use the SDSS photometry reported by \citet{2020ApJS..250....8L}, and for eFEDS-selected sources in SDSS DR17 we use the Hyper-Suprime Cam photometry reported by \citet{2022A&A...661A...3S}. 
Eddington luminosities are calculated in the usual way, balancing the gravitational and radiation forces in a Hydrogen-only plasma, and assuming the dominant source of opacity is Thomson electron scattering:
\begin{equation}
    L_\textrm{Edd} = \frac{4\pi G M_\textrm{BH} m_\textrm{p} c}{\sigma_\textrm{T}} = 1.26 \times 10^{38} \left(\frac{M_\textrm{BH}}{M_\odot}\right) \ergps.
\end{equation}
The Eddington ratio $L_\textrm{bol}/L_\textrm{Edd}$ (hereafter $L/L_\textrm{Edd}$) is then estimated assuming a constant bolometric correction of $L_\textrm{bol} = 5.15 \times L_{3000}$. We discuss this assumption further in Section~\ref{sec:bolometric_corr}, and show how our key observables depend directly on FWHM(\mgii) and $L_{3000}$ in Appendix~\ref{appendixd}.

Our sample of  186\,303 quasars spans 2.5\,dex in luminosity, with $L_{3000} \approx 10^{44.5-47}\ergps$ and $L_\textrm{bol} \approx 10^{45-47.5}\ergps$.
We infer SMBH masses in the $10^{8-10} M_\odot$ range 
and Eddington ratios from 0.01 to unity, with the distribution of our sample shown in Fig.~\ref{fig:sample}.

\section{Modeling the Quasar SED}
\label{sec:models}

Our second goal is to confront observational data with models for accretion and outflow in quasars; more specifically, we aim to test if the changes in observed emission line and continuum properties with $M_\textrm{BH}$ and Eddington ratio are consistent with
\qsosed\footnote{
\url{https://heasarc.gsfc.nasa.gov/xanadu/xspec/manual/node132.html}} 
models \citep{2018MNRAS.480.1247K}
for the SED of the ionizing continuum.
We used the implementation of \qsosed\ in \textsc{xspec} \citep{xspec} to calculate SEDs, via the \textsc{pyxspec} python wrapper \citep{pyxspec}. 

In \qsosed, the radiation originates from three characteristic regions: an outer thermal disc, an inner hot Comptonising `corona' and an intermediate warm Comptonising component. These three regions are assumed to be radially stratified as defined by four critical radii: $R_\textrm{ISCO} < R_\textrm{hot} < R_\textrm{warm} < R_\textrm{out}$.
The inner and outer radii are defined by the radius of the innermost stable circular orbit $R_\textrm{ISCO}$ and the self-gravitation radius $R_\textrm{out}$.
The hot X-ray component originates from $R_\textrm{ISCO}<R<R_\textrm{hot}$, and has a luminosity set by the sum of the directly dissipated power, $L_\textrm{diss,hot}$, and the seed photon luminosity, $L_\textrm{seed}$. One of the key aspects of the model is the empirically motivated assumption that the dissipated power is always $2$ per cent of the Eddington luminosity; this constraint defines the value of $R_\textrm{hot}$. 
The outer radius of the warm Comptonising component $R_\textrm{warm}$ is set to be twice $R_\textrm{hot}$. For $R_\textrm{warm}<R<R_\textrm{out}$, the thermal disc component is assumed to emit as described by \citet{1973blho.conf..343N}.

\qsosed\ has four physical input parameters: the cosine of the inclination, $\cos i$, the SMBH mass, $M_\textrm{BH}$, the dimensionless spin parameter, $a_*$, and the Eddington-scaled accretion rate, $\dot{m} \equiv \dot{M}_\textrm{acc}/\dot{M}_\textrm{Edd}$. We fix $\cos i = 0.5$ and calculate grids of models in $(M_\textrm{BH},\dot{m})$ parameter space, for non- and maximally-spinning SMBHs $a_* \in (0,0.998)$. We calculate models with 21 logarithmically-spaced grid points in each direction, spanning the ranges $8 \leq \log (M_\textrm{BH}/M_\odot) \leq 10$ and $-1.65 \leq \log \dot{m} \leq 0$, corresponding to intervals of $0.1$ and $0.0825$~dex. To compare with observations, we take the input SMBH mass and calculate the Eddington ratio from $L_{3000}$ using the same bolometric correction of $5.15$ that we apply to the observational data (but see Section~\ref{sec:bolometric_corr} and Appendix~\ref{appendixb}). Here, and in Section~\ref{sec:results:models}, we present models for only the non-spinning case, as these are in much better agreement with the data. We discuss the impact of SMBH spin and system inclination in Section~\ref{sec:spin_inc} and models with $a_*=0.998$ are presented in Appendix~\ref{appendixc}. 

Although the emission line properties must depend on the ionizing SED, the exact relationship between, for example, \civ\ EW and the SED is complex due to a number of confounding factors such as BLR geometry, density and radiative transfer. The relationship to any kinematic signatures such as \civ\ blueshift is even more complicated and would require a physical model for the line formation region and associated flow dynamics. A somewhat simpler case is the EW of \ion{He}{II}\,$\lambda$1640, which is a recombination line and therefore a reasonable `photon counter'. \ion{He}{II} has history as a tracer of the EUV continuum: for example, in cataclysmic variables 
\cite{hoare1991} applied a modified  \citet{zanstra1929} method to infer boundary layer temperatures, and in quasars, \cite{2004ApJ...611..125L} note that a high \ion{He}{II} EW is indicative of a strong X-ray continuum. Assuming Case B recombination, \cite{1987ApJ...323..456M} give the \ion{He}{II}\,$\lambda$1640 EW in terms of the 228\,\AA\ continuum flux. Their equation can be inverted to give the proportionality
\begin{equation}
     \frac{F_\nu (\lambda 228)}{F_\nu (\lambda 1640)} \propto \textrm{EW} ({\rm\ion{He}{II}~\lambda1640}) \frac{\Omega}{4\pi},
\end{equation}
where ${\Omega}/{4\pi}$ is the covering fraction and the proportionality constant is dependent on the shape of the SED (\citealt{1987ApJ...323..456M} considered a power law in $F_\nu$ at $228\,\textrm{\AA}$). 
In this work we assume, based on the above equation, that the observed \ion{He}{II}\,$\lambda$1640 EW is a reasonable proxy for the ratio of continuum luminosities $L_{228}/L_{1640}$.

\begin{figure}
	\includegraphics[width=\columnwidth]{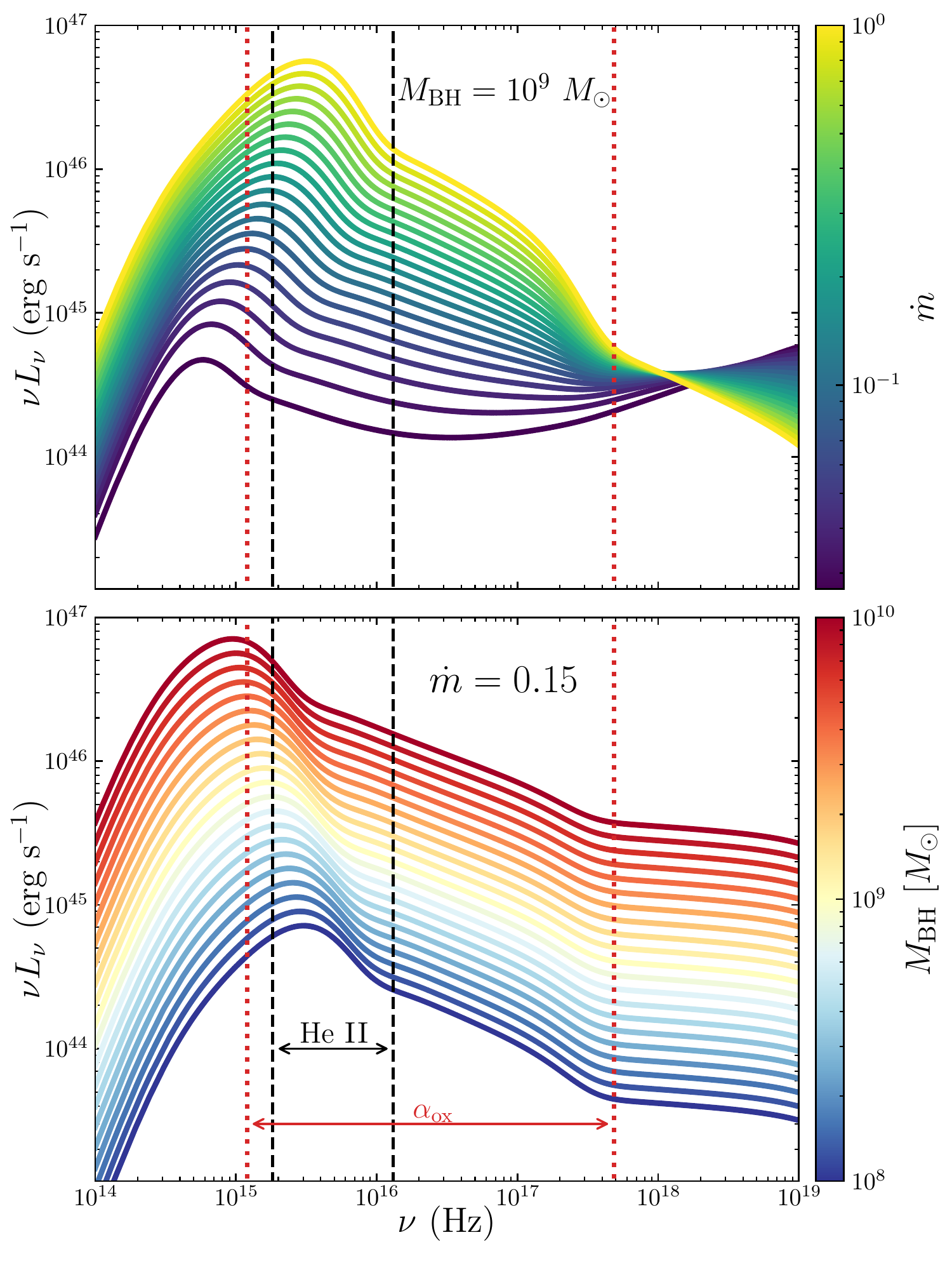}
    \caption{Output SEDs in $\nu L_\nu$ units from \qsosed\ for $a_*=0$ and $\cos i =0.5$. 
    The vertical lines show, from left to right, the frequencies at 2500\,\AA, 1640\,\AA, 228\,\AA\ (=\,54\,eV) and 2\,keV which together determine $\alpha_\textrm{ox}$ and the EW of \ion{He}{II}\,$\lambda$1640. The \ion{He}{II} ionization edge at 54\,eV ($1.3\times 10^{16}$\,Hz) lies in the EUV regime where the intermediate warm Comptonising component in \qsosed\ is most important, but the EW of \ion{He}{II} can also be seen to depend on the location of the peak of the ionizing SED.
    \textsl{Top panel:} SEDs with fixed SMBH mass of $10^9\,M_\odot$ and varying $\dot{m}$ in logarithmic intervals. As $\dot{m}$ increases the peak of the SED moves to the blue, the luminosity increases, and the hard X-ray power law spectral index becomes softer.
    \textsl{Bottom panel:} SEDs with fixed $\dot{m}=0.15$ and varying $M_\textrm{BH}$ in logarithmic intervals. As SMBH mass increases the peak of the SED  moves to the red, and the luminosity increases. A maximal spin analogue to this plot is shown in Fig.~\ref{fig:bh_spin}. 
    }
    \label{fig:qso_seds}
\end{figure}

In Fig.~\ref{fig:qso_seds} we present output SEDs  from \qsosed, in which the three radially stratified components can be seen as separate `bumps' in the spectrum. In these plots, we show how the  model SEDs change as a function of Eddington-scaled accretion rate, $\dot{m}$ (for fixed mass, top panel) and SMBH mass, $M_\textrm{BH}$ (for fixed $\dot{m}$, bottom panel). The important frequencies for determining \ion{He}{II} EW (corresponding to $228$\,\AA\ and $1640$\,\AA)  and $\alpha_\textrm{ox}$ (corresponding to $2500$\,\AA\ and $2$~keV) are marked. Increasing $\dot{m}$ increases the overall luminosity of the system and pushes the peak of the outer thermal disc component to higher frequencies. Simultaneously, the hard X-ray slope becomes significantly softer and $L_{2 \textrm{keV}}$ only increases slowly,
consistent with eq.~6 and fig.~5b of \citet{2018MNRAS.480.1247K}.
As a result, the higher Eddington fraction objects are more X-ray weak relative to their ultraviolet flux. Increasing $M_\textrm{BH}$ also increases the total luminosity, but now the peak of the thermal component moves to lower frequencies and the hard X-ray slope  stays fairly constant. In both panels of the plot the peak of the SED can be found on either side of the low frequency pivot points for both \ion{He}{II} EW and $\alpha_\textrm{ox}$, resulting in an interesting interplay between these quantities and the fundamental AGN parameters.

\section{Results}
\label{sec:results}

\begin{figure*}
	\includegraphics[width=\columnwidth]{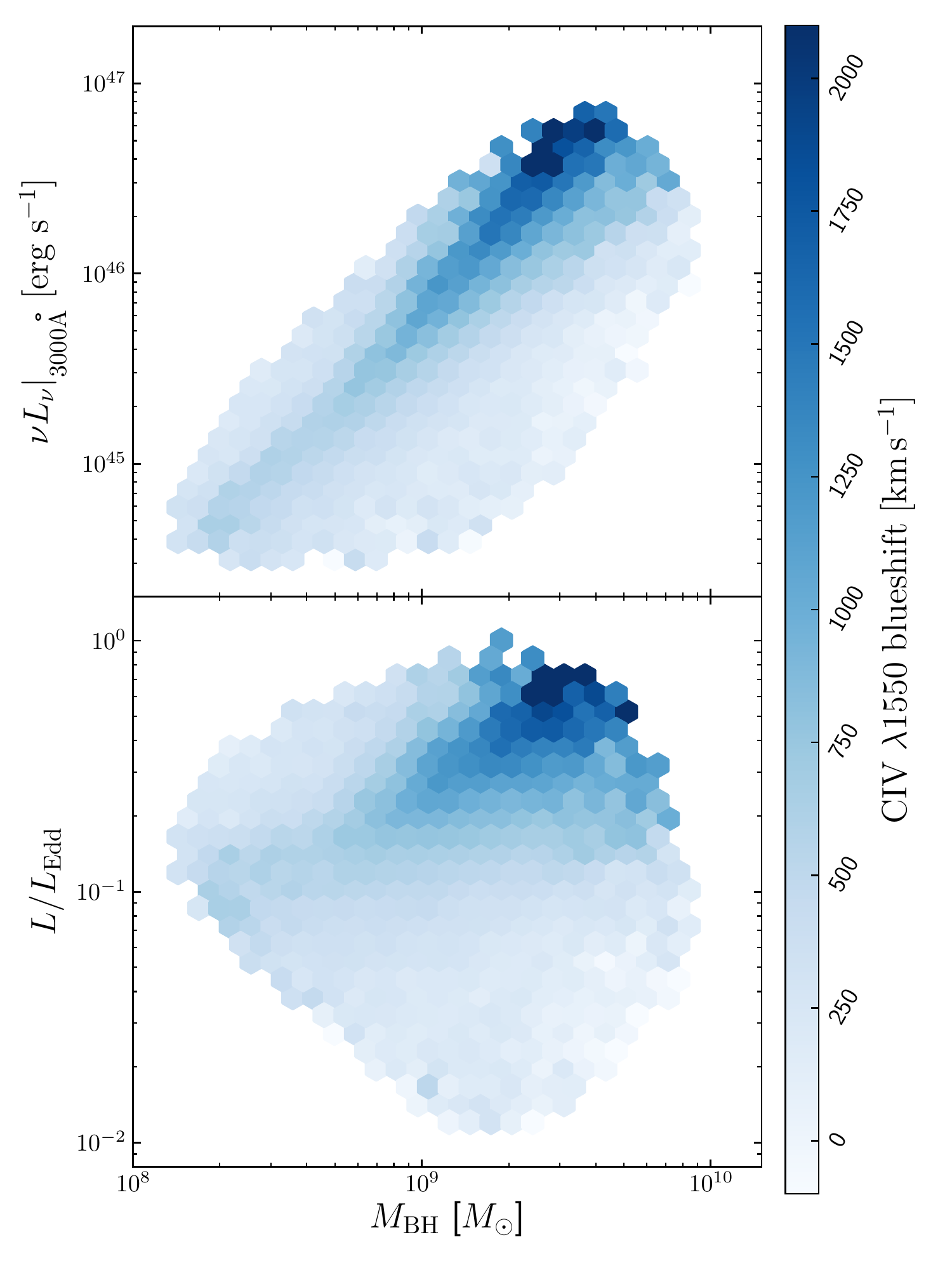}
	\includegraphics[width=\columnwidth]{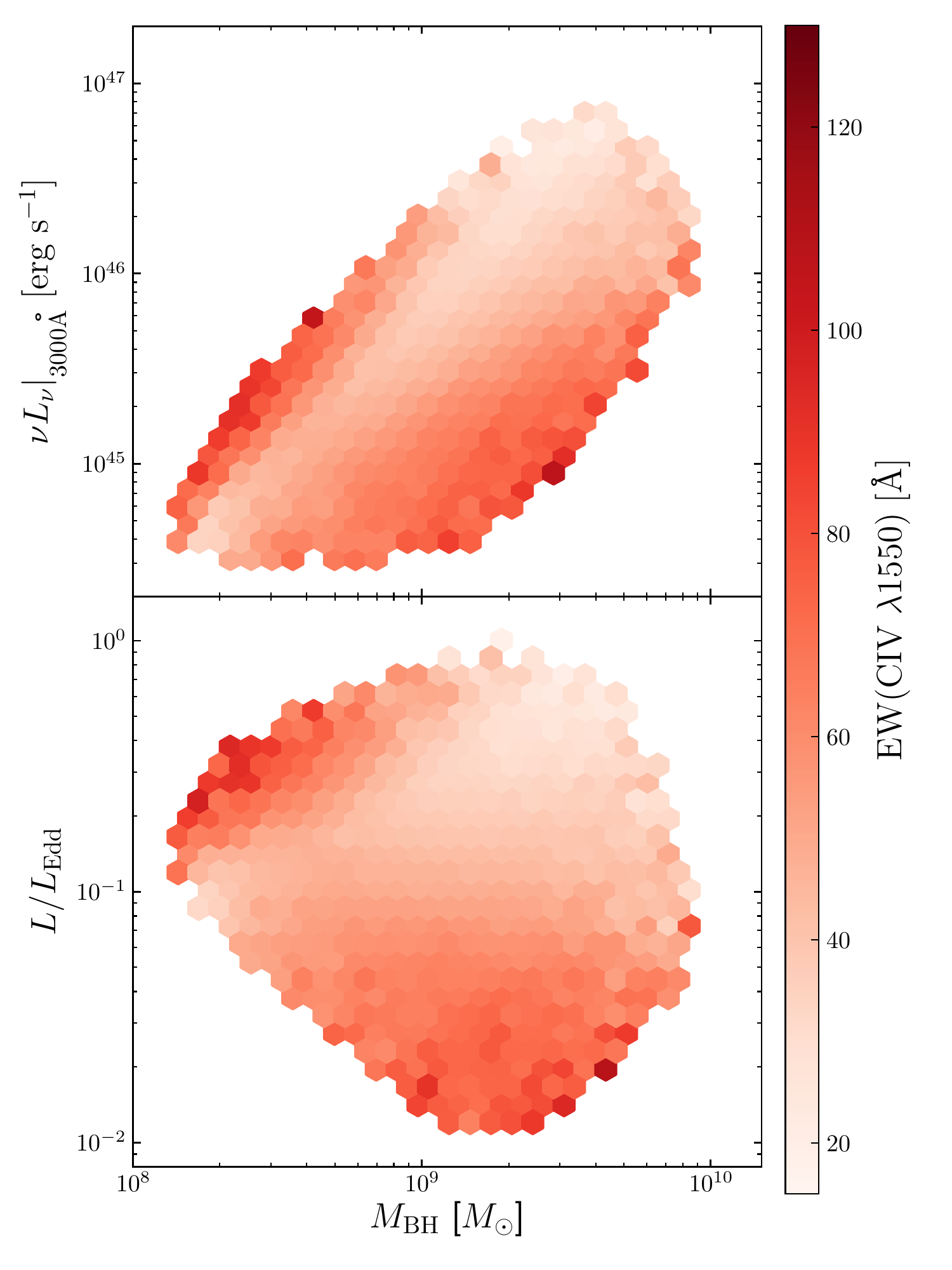}
    \caption{%This is an example figure. Captions appear below each figure.
	The median observed \ion{C}{IV} blueshift (left) and EW (right) in bins of SMBH mass, 3000\,\AA\ ultraviolet continuum luminosity (top) and Eddington ratio (bottom).	Data are shown only for bins which contain five or more objects.
    The \civ\  blueshift and EW are seen to anti-correlate: areas of parameter space with strong blueshifts have weak EW and vice versa.
    $L/L_\textrm{Edd}\gtrsim0.1$ is a necessary but not sufficient condition for observing the largest \civ\ blueshifts. 
    The strongest \civ\ blueshifts are observed only at large SMBH mass and large Eddington ratio, while high EW \civ\ emission is observed at large Eddington ratio and smaller mass. 
    The Baldwin effect can be observed in the sense that objects with brighter 3000\,\AA\ luminosities tend to have weaker \civ\ EWs on average. However, the \civ\ EW behaviour as a function of $M_\textrm{BH}$ and $L/L_\textrm{Edd}$ shows that the underlying drivers of the Baldwin effect are more complicated than a simple dependence on the ultraviolet luminosity.
	}
    \label{fig:CIV}
\end{figure*}

\begin{figure*}
	\includegraphics[width=\columnwidth]{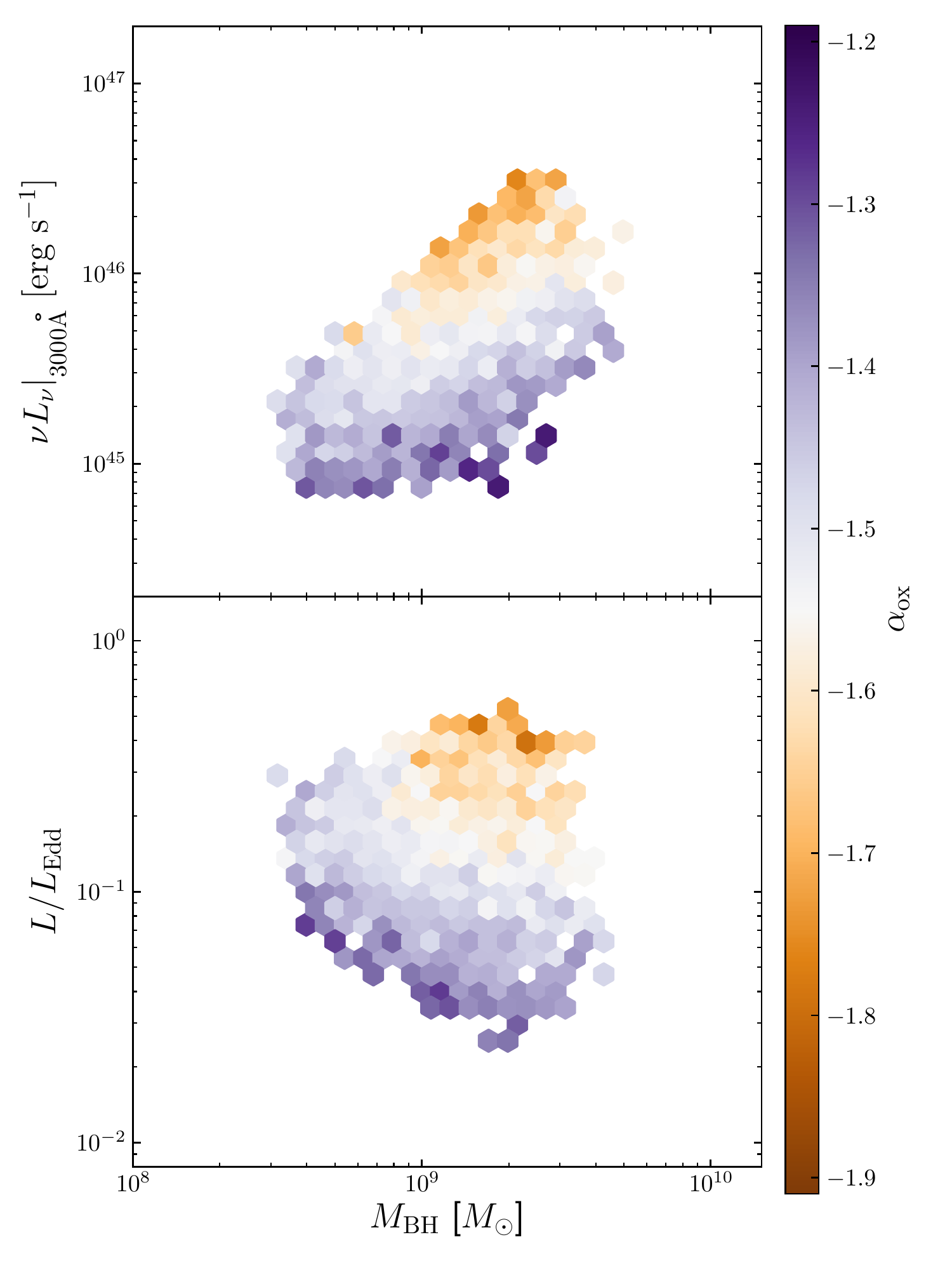}
	\includegraphics[width=\columnwidth]{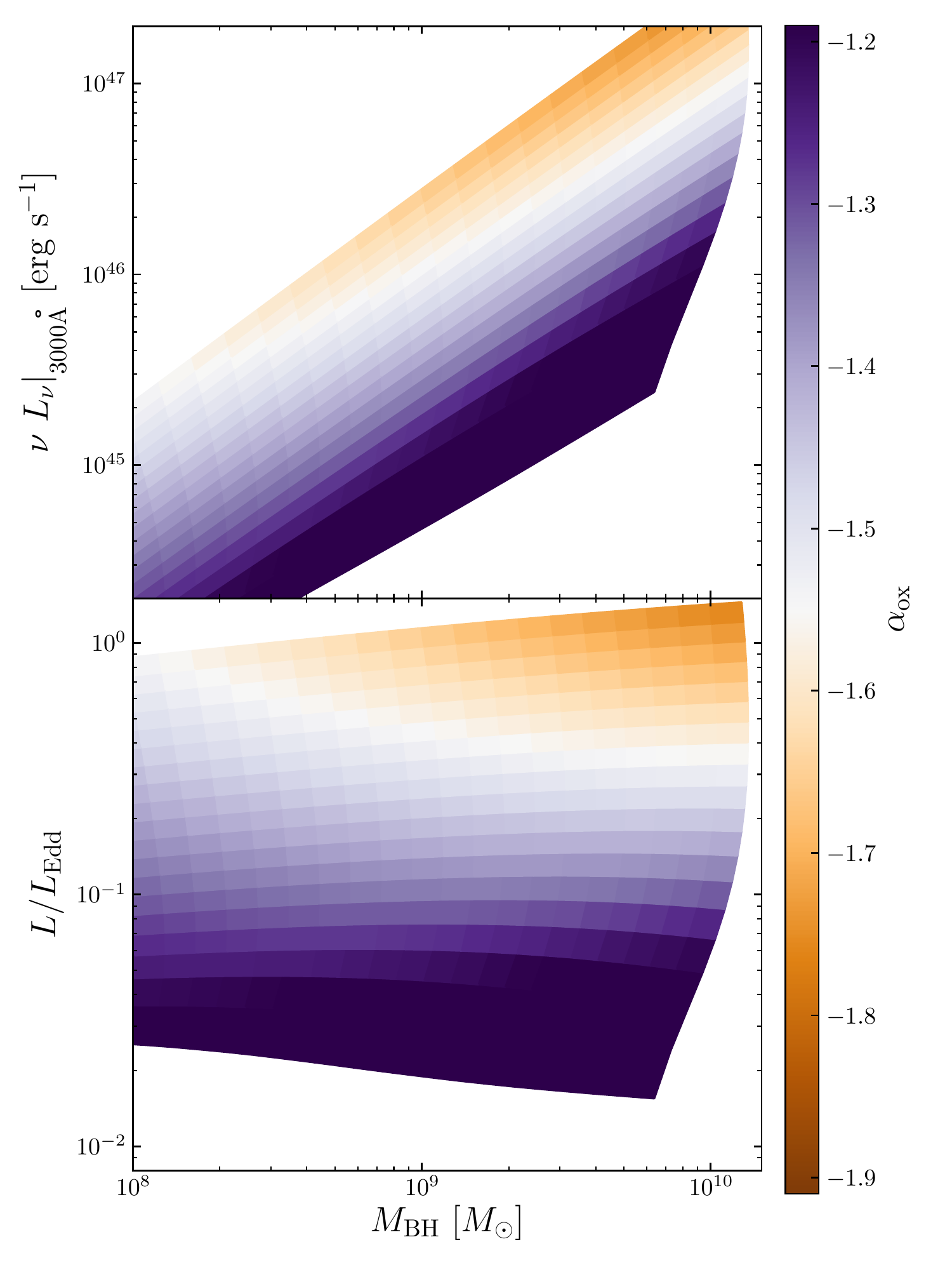}
    \caption{
    \textsl{Left panel:} The median observed $\alpha_\textrm{ox}$ in bins of SMBH mass, 3000\,\AA\ ultraviolet continuum luminosity (top) and Eddington ratio (bottom) for the 5031 objects from our sample with 2\,keV X-ray measurements. Data are shown only for bins which contain five or more objects.
    \textsl{Right panel:} the predicted $\alpha_\textrm{ox}$ from low spin \qsosed\ models in the same parameter space.
    The observations show good agreement with the models, with $\alpha_\textrm{ox}$ more negative (i.e.\,more X-ray weak) in objects with brighter ultraviolet luminosities.
    In Fig.~\ref{fig:bh_spin} we show equivalent models but with high spin, which do not show such agreement with the observations, suggesting that the $z\approx2$ SDSS quasar population may be more consistent with low SMBH spins on average.
    }
    \label{fig:aox}
\end{figure*}

\begin{figure*}
	\includegraphics[width=\columnwidth]{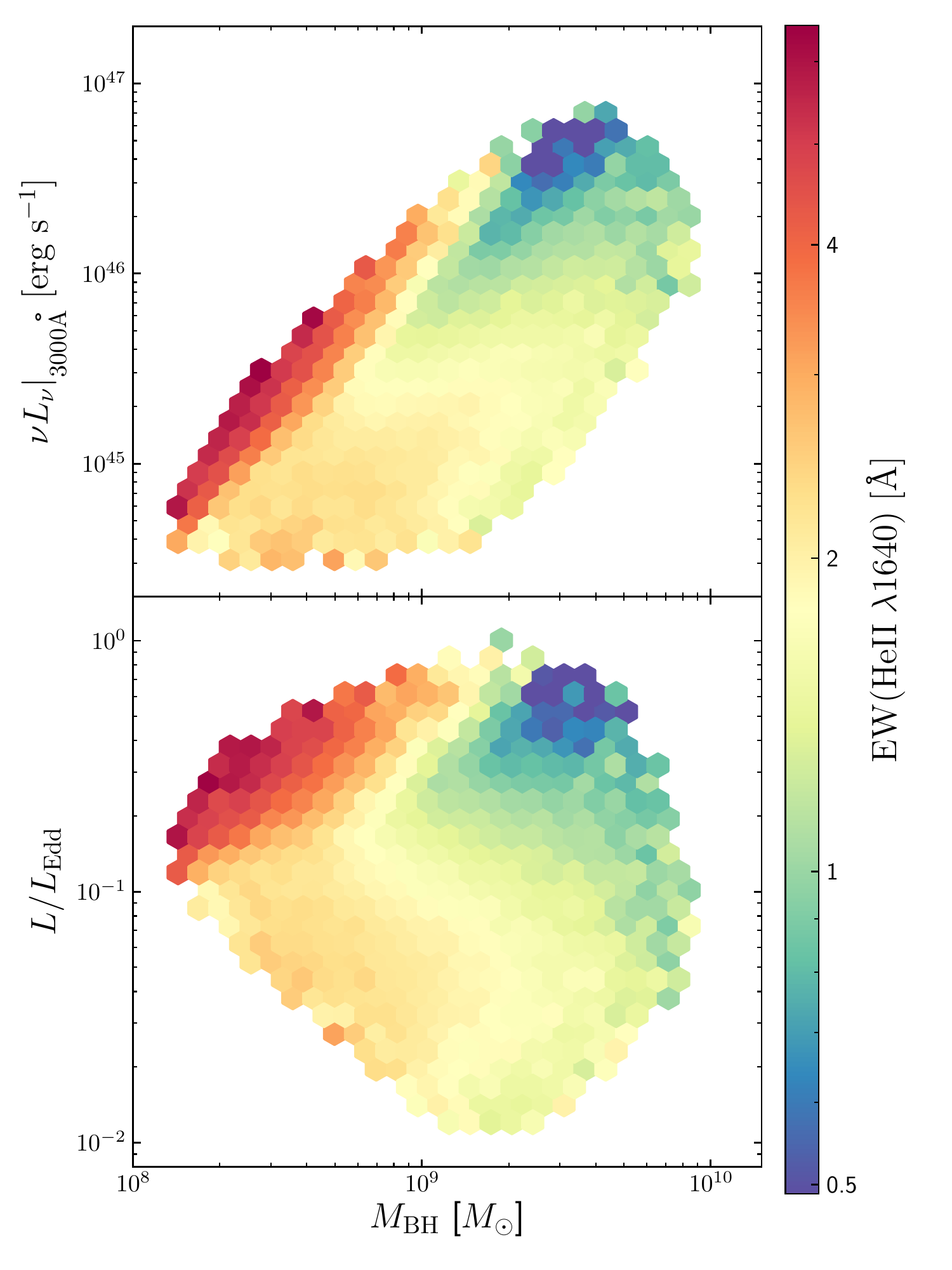}
	\includegraphics[width=\columnwidth]{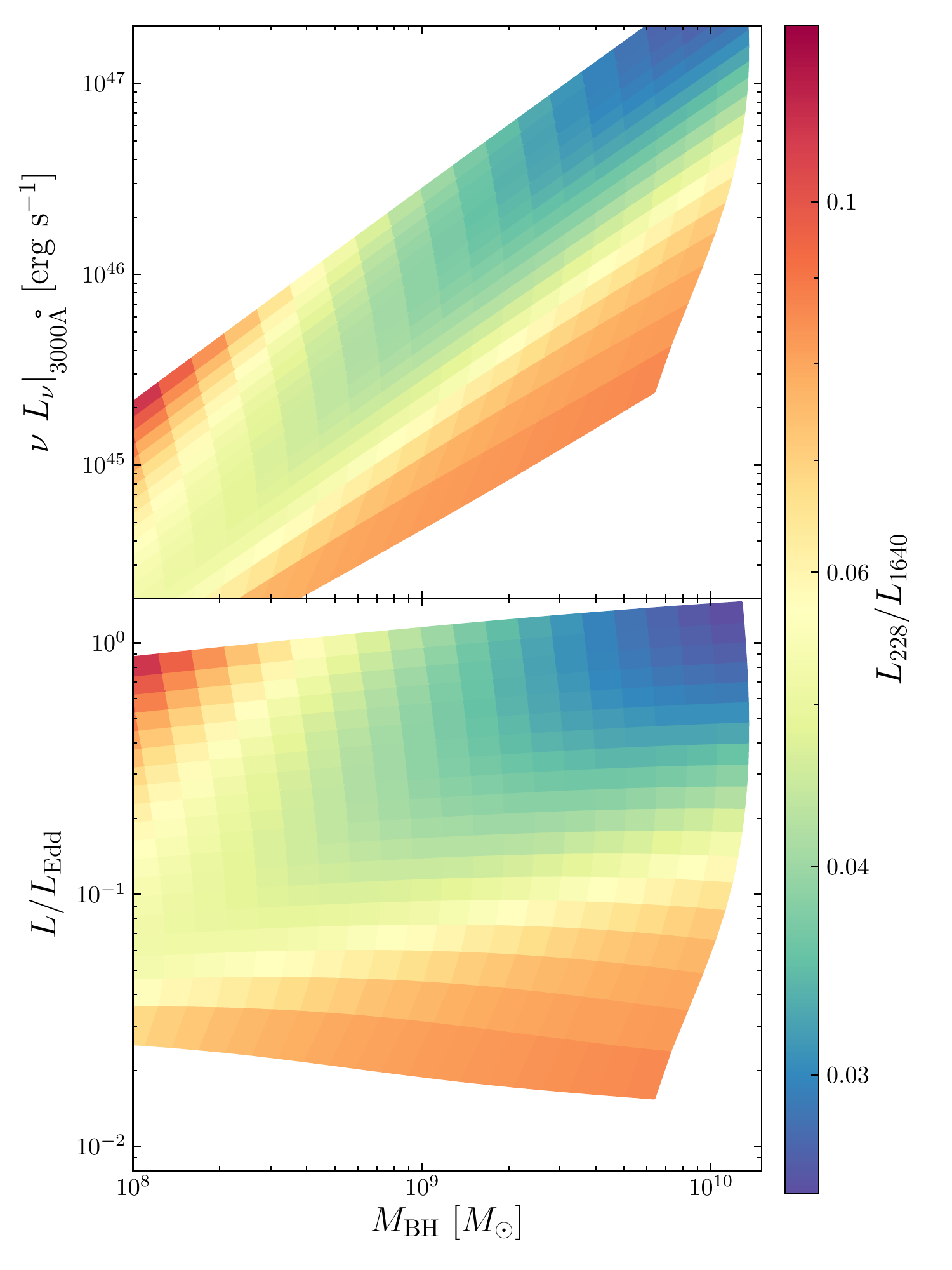}
    \caption{
    \textsl{Left panel:} The median observed \ion{He}{II} EW in bins of SMBH mass, 3000\,\AA\ ultraviolet continuum luminosity (top) and Eddington ratio (bottom).	Data are shown only for bins which contain five or more objects.
    \textsl{Right panel:} the predicted strength of \ion{He}{II} ionizing photons at 228\,\AA\ relative to the 1640\,\AA\ continuum from \qsosed\ models.
    Above an Eddington ratio of $\approx$0.1,
    there is a strong trend as a function of SMBH mass, with high mass objects showing the weakest \ion{He}{II} emission and low mass objects showing the strongest \ion{He}{II} emission. 
    The model predictions show qualitatively similar behaviour in this region of parameter space, explaining the diagonal contours in constant \ion{He}{II}.
    Below $L/L_\textrm{Edd}\lesssim0.1$,
    the observed \ion{He}{II} displays much weaker trends, and does not agree with the model predictions, suggesting that in this regime either the SED models are less accurate or the structure of the BLR is changing.
    }
    \label{fig:HeII}
\end{figure*}

\subsection{Observed properties in $M_\textrm{BH}$--$L/L_\textrm{Edd}$ space}
\label{sec:results:obs}

The first observational result from this work is the behaviour of the \ion{C}{IV}\,$\lambda$1549 emission line morphology as a function of SMBH mass ($M_\textrm{BH}$) and Eddington ratio ($L/L_\textrm{Edd}$), shown in Fig.~\ref{fig:CIV}.
In the left panel, we show the \ion{C}{IV} emission line blueshift (as defined in Eq.~\ref{eq:blue}) and in the right panel the EW of \ion{C}{IV}. In the top panels, consistent with previous works, we find that more luminous quasars show weaker emission line strengths relative to the continuum and stronger emission line blueshifts. However, when considering the observed \civ\ properties as a function of both $M_\textrm{BH}$ and $L/L_\textrm{Edd}$ (bottom panel), we see a more complicated behaviour.
To observe the strongest \ion{C}{IV} blueshifts (which are associated with the smallest EWs), we need to look at objects with both $M_\textrm{BH} \gtrsim 10^9 M_\odot$ and $L/L_\textrm{Edd} \gtrsim 0.1$. Moreover, the contours of constant \ion{C}{IV} blueshift follow acute-angled `wedge' shapes, which are somewhat orthogonal to lines of constant luminosity (running diagonally top-left to bottom-right in the $M_\textrm{BH}$--$L/L_\textrm{Edd}$ space).
At the same time, objects with the strongest \ion{C}{IV} EWs $\gtrsim 100$\,\AA, which have strong symmetric emission with little or no blueshift, are found at high $L/L_\textrm{Edd} \gtrsim 0.1$ and relatively low $M_\textrm{BH} \lesssim 10^9 M_\odot$,
and also at low $L/L_\textrm{Edd} \lesssim 0.03$.

To help us to understand the physical drivers behind the trends seen in \ion{C}{IV}, in Figs.~\ref{fig:aox} and \ref{fig:HeII} we also show  $\alpha_\textrm{ox}$ and  \ion{He}{II} EW across the same $M_\textrm{BH}$--$L/L_\textrm{Edd}$ parameter space.
The $\alpha_\textrm{ox}$ behaviour is as expected from previous works \citep[e.g.][]{2022arXiv221011977M}, largely with more luminous objects displaying relatively weaker X-ray emission which is quantified by a more negative $\alpha_\textrm{ox}$.
A more interesting result is seen in the EW of \ion{He}{II}, which is even more striking than the behaviour seen in \ion{C}{IV}. With the \ion{He}{II} EW, there is a clear transition around $L/L_\textrm{Edd} \approx 0.1$, with both the strongest and weakest line emission only seen above this threshold.
Below this Eddington limit, there is little change in the average line properties as a function of mass, but at $L/L_\textrm{Edd} \gtrsim 0.1$ there is a strong mass dependence with diagonal wedge-shaped contours similar to those observed in \ion{C}{IV}.
By contrast, the contours of constant $\alpha_\textrm{ox}$ are much less closely aligned with contours of constant \civ\ blueshift.

To test the robustness of these trends, we divide the $M_\textrm{BH}$--$L/L_\textrm{Edd}$ into square bins of 0.1 by 0.1\,dex and compute the median absolute deviation (MAD) in each bin. 
% approx 400 objects per bin on average
The typical MAD is 285\,\kmps\ in \civ\ blueshift, 13\,\AA\ in \civ\ EW and 0.49\,\AA\ in \ion{He}{II} EW. The typical scatter within each bin is therefore significantly less than the dynamic range in the average emission line properties shown in Figs.~\ref{fig:CIV} and \ref{fig:HeII}, meaning that one is unlikely to find individual objects which go against the overall trend of the population.
Dividing through by the median in each bin, the typical MAD/median in each bin is 0.29 and 0.26 for the \ion{He}{II} and \civ\ EWs respectively, meaning that the typical range of emission line EW within each $M_\textrm{BH}$--$L/L_\textrm{Edd}$ bin is a factor of 3.5 and 3.8 for \ion{He}{II} and \civ\ respectively, compared with the dynamic range of more than a factor of six seen in the median per-bin line properties.

\subsection{Comparison with model SEDs}
\label{sec:results:models}

In the right-hand panels of Figs.~\ref{fig:aox} and \ref{fig:HeII} we show how $\alpha_\textrm{ox}$ and 
$L_{228}/L_{1640}$, respectively, vary with mass and Eddington fraction, as modeled by \qsosed. These plots can be compared to the respective plots from the observational sample (left-hand panels), albeit with some caveats regarding bolometric corrections (Section~\ref{sec:bolometric_corr}) and $M_\textrm{BH}$ estimates (Section~\ref{sec:assumptions}). In a qualitative sense, the models do a reasonably good job of reproducing the trends observed in the data. Focusing first on $\alpha_\textrm{ox}$, we can see that the general trend of decreasing $\alpha_\textrm{ox}$ with Eddington fraction is reproduced, and, in addition, the gradient is stronger at high $M_\textrm{BH}$, as observed in the data. To put this another way, in both the data and model results, the contour of fixed $\alpha_\textrm{ox}$ curves around, from being nearly horizontal at high $M_\textrm{BH}$ to being closer to vertical at low $M_\textrm{BH}$. The dynamic range of model $\alpha_\textrm{ox}$ values is comparable to that observed, but the models do not produce soft enough spectra to match the data; $\alpha_\textrm{ox} \approx -1.9$ can be found in some bins in the quasar sample but the minimum value of $\alpha_\textrm{ox}$ in the models is $-1.79$. 

The comparison of the model $L_{228}/L_{1640}$ ratio and the observed \ion{He}{II}\,$\lambda1640$ EW is also broadly encouraging, at least at relatively high Eddington fractions. This finding is perhaps more interesting as the \ion{He}{II} EW is probing a portion of the SED that is not accessible directly. The basic behaviour, of decreasing \ion{He}{II} EW with $M_\textrm{BH}$ at high Eddington fractions, is well matched by the models. The models also capture the diagonal contours of constant \ion{He}{II} EW, in which the transition to low \ion{He}{II} EWs occurs at higher masses for higher Eddington fractions. As discussed above, at low Eddington fractions ($L/L_\textrm{Edd} \lesssim 0.1$), something fundamentally switches in the data, with gradients generally being shallower and along a different direction in the parameter space. This relatively sharp change is not  reproduced by the models, and may be telling us something fundamental about the quasar accretion process (see Section~\ref{sec:discuss_accretion} for a discussion).

\section{Discussion}
\label{sec:discuss}

We have quantified the average behaviour of \ion{C}{IV}\,$\lambda 1549$, \ion{He}{II}\,$\lambda1640$ and $\alpha_\textrm{ox}$ as a function of both  $M_\textrm{BH}$ and $L/L_\textrm{Edd}$, and compared our observations with predictions from \qsosed\ models. In this section we now discuss these results.
We first outline the key caveats in our findings (Sections~\ref{sec:bolometric_corr} and \ref{sec:assumptions}), before discussing possible interpretations of our results within the context of AGN accretion and outflow theories (Sections~\ref{sec:discuss_accretion} and \ref{sec:discuss_outflows}).
Finally, we discuss some wider implications and possible future applications (Section~\ref{sec:discuss:future}), before summarizing our key conclusions in Section~\ref{sec:conclude}.

\subsection{Key assumptions and limitations}
\subsubsection{Bolometric corrections}
\label{sec:bolometric_corr}

\begin{figure}
	\includegraphics[width=\columnwidth]{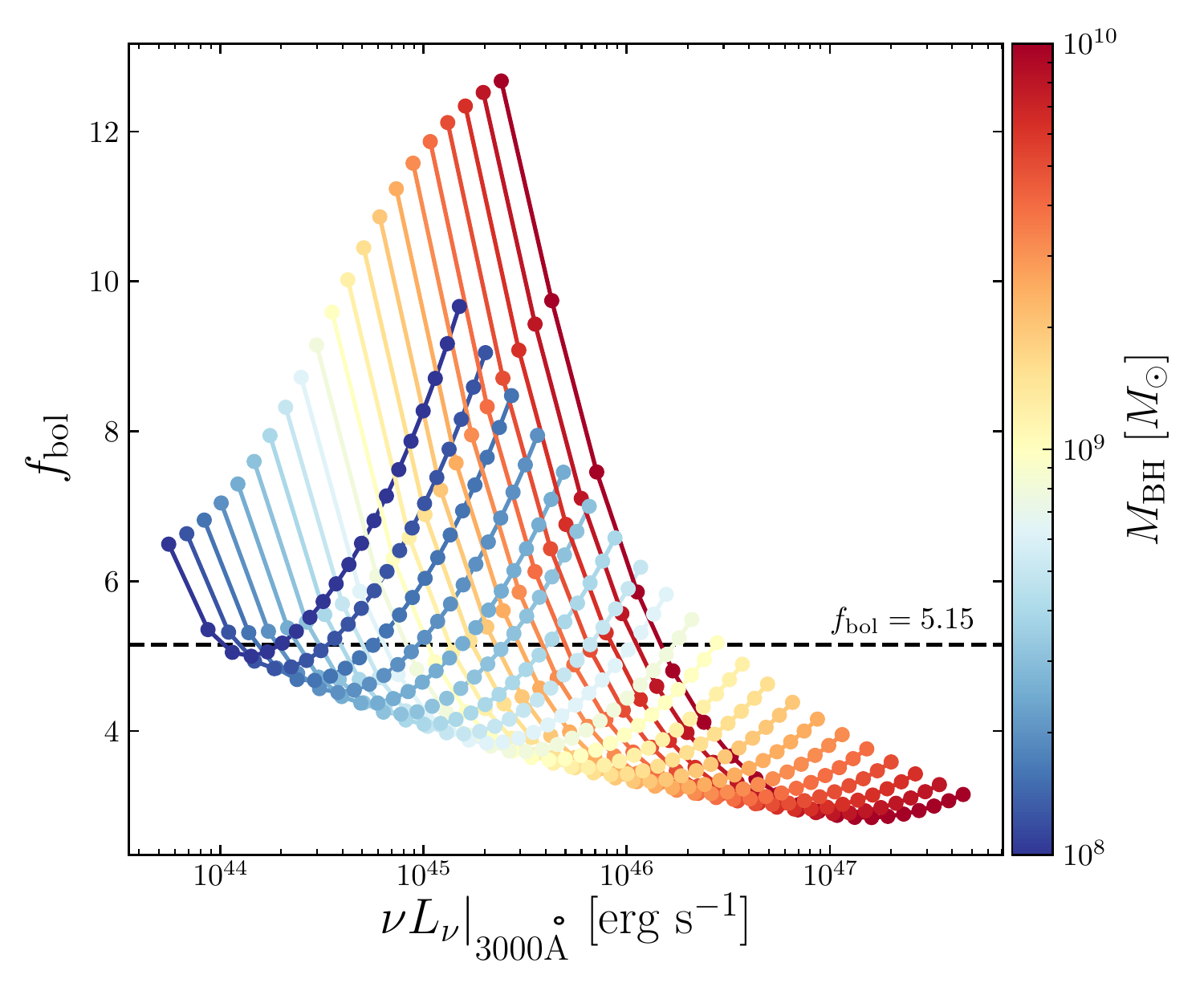}
    \caption{The predicted bolometric correction, $f_\textrm{bol} \equiv  L_\textrm{bol} / L_{3000}$, as a function of $L_{3000}$, from \qsosed\ models. The points are colour-coded by $M_\textrm{BH}$ with a logarithmic normalisation, and points of constant mass are joined with solid lines so that the trends with Eddington ratio can be understood by following individual lines from left to right. The adopted bolometric correction in this work, $f_\textrm{bol}=5.15$, is shown as a horizontal dashed line. $f_\textrm{bol}$ ranges from $\approx 3-10$, and our adopted $f_\textrm{bol}$ is bounded by this range; however $f_\textrm{bol}$ does have a clear dependence on mass and luminosity in the model SEDs.
    Our assumption of a fixed $f_\textrm{bol}$ could lead to an artificially reduced dynamic range in the inferred $L_\textrm{bol}$ at $M_\textrm{BH}=10^8M_\odot$ and an artificially increased range of $L_\textrm{bol}$ at $M_\textrm{BH}=10^{10}M_\odot$. }
    \label{fig:bolometric}
\end{figure}

A large part of this work has attempted to quantify the `unseen' extreme ultraviolet (EUV) portion of the SED which is not directly observable, but which can instead be probed via the \ion{He}{II} emission line. This portion of the SED contributes a significant amount to the bolometric luminosity of a quasar. To estimate bolometric luminosities (and Eddington ratios $L_\textrm{bol} / L_\textrm{Edd}$), we have assumed a constant bolometric correction $f_\textrm{bol} \equiv  L_\textrm{bol} / L_{3000}$ of 5.15, consistent with previous works in the literature \citep[e.g.][]{2006ApJS..166..470R, 2013ApJS..206....4K}. However, we have also shown that the \ion{He}{II} strength is changing as a function of $M_\textrm{BH}$ and $L/L_\textrm{Edd}$, so we expect the strength of the EUV continuum and hence the bolometric correction to be varying with $M_\textrm{BH}$ and $L/L_\textrm{Edd}$.
Using our \qsosed\ models, we  attempt to quantify this effect in Fig.~\ref{fig:bolometric}. While our chosen value of  $f_\textrm{bol}=5.15$ lies within the range of values spanned by our grid of model SEDs, there is  variation of around a factor of two in  $f_\textrm{bol}$ depending on the values of $M_\textrm{BH}$ and $L/L_\textrm{Edd}$ we consider. While this could in principle lead to systematic biases in our estimation of $L/L_\textrm{Edd}$, we show in Appendix~\ref{appendixb} that these biases are likely to be small compared to the magnitude of the trends we observe.

We can however, briefly describe what might happen if we were to adopt a non-constant bolometric correction when inferring $L/L_\textrm{Edd}$ from our observations. 
For two objects, both at $\dot{m}=0.2$, the $f_\textrm{bol}$ inferred from the \qsosed\ models would be $\approx$6 and $\approx$3 for $M_\textrm{BH}=10^{8}M_\odot$ and $10^{10}M_\odot$ respectively.
This would skew the observations in Fig.~\ref{fig:HeII}, moving the location of the strongest \ion{He}{II} EW (at low $M_\textrm{BH}$) to larger $L/L_\textrm{Edd}$, more in line with the $L/L_\textrm{Edd}$ threshold at high $M_\textrm{BH}$ above which we see the weakest \ion{He}{II} and largest \civ\ blueshifts.

\subsubsection{Black hole mass estimates}
\label{sec:assumptions}

As well as the assumption of a constant bolometric correction, we have used a single-epoch virial estimator to estimate SMBH masses throughout this work.
The caveats associated with such estimates are numerous and have been reviewed by  \citet{2013BASI...41...61S}. Here we discuss some of the issues which are most relevant to our method and results. Most notably, the BLR radius--luminosity relation (as encoded through the virial $f$ factor) may depend on the shape of the SED. Other uncertainties arising from (for example) orientation are likely to be random, in the sense that they will add scatter to our $M_\textrm{BH}$ estimates but should not bias our results.
While it is  possible that our observed distribution of quasars in the $M_\textrm{BH}$--$L/L_\textrm{Edd}$ plane is not the same as the intrinsic distribution, 
the fact that we do still observe such striking behaviour in the \ion{He}{II} and \ion{C}{IV} emission line properties as a function of our inferred $M_\textrm{BH}$ and $L/L_\textrm{Edd}$ is telling us that any random scatter or noise in our $M_\textrm{BH}$ estimates is small enough not to `wash out' the observed trends.

We  used  the FWHM of the \ion{Mg}{II} line to estimate $M_\textrm{BH}$.
\citet{2008ApJ...680..169S} showed that such \ion{Mg}{II}-derived $M_\textrm{BH}$ estimates correlate tightly with those derived from H$\beta$ across the full $10^{8-10}\,M_\odot$ mass range, with the distribution of $\textrm{log}\left(M_\textrm{BH}^{\textrm{H}\beta}\big/ M_\textrm{BH}^{\ion{Mg}{II}}\right)$ following a Gaussian with mean 0.034 and dispersion 0.22\,dex.
\citet{2012ApJ...753..125S}  extended this analysis to higher redshifts and higher luminosities, more appropriate for the objects in this work, and again found that the \ion{Mg}{II} properties remained well correlated with those of $\textrm{H}\beta$.
The \ion{Mg}{II}-derived $M_\textrm{BH}$ estimates we use in this work are therefore unlikely to be biased compared to those which we would have derived from a single-epoch H$\beta$ measurement. The possibility remains, however, that such estimates are biased as a function of the SED, or equivalently, as a function of $M_\textrm{BH}$ and $L/L_\textrm{Edd}$.

Early concerns about the universality of the BLR radius--luminosity relation were discussed by \citet{2005ApJ...629...61K} and \citet{2006A&A...456...75C}.
More recently, various authors have tried to account for possible SED-dependent biases in single-epoch $M_\textrm{BH}$ estimates \citep{2019ApJ...886...42D, 2020ApJ...903..112D, 2020ApJ...899...73F, 2020ApJ...903...86M}, either using the  accretion rate directly or by using the strength of optical iron emission $R_\ion{Fe}{II}$ as a proxy.
However, \citet{2022MNRAS.513.1985K, 2022MNRAS.515.3729K} and \citet{2022arXiv220805491Y} find the opposite result, with the inclusion of $R_\ion{Fe}{II}$ having no effect on the scatter in either the  \ion{Mg}{II} or H$\beta$ radius--luminosity relations.

While the literature is divided, we argue it is still true that any SED-dependent bias in our single-epoch $M_\textrm{BH}$ estimates must be contained within the scatter on the BLR radius--luminosity relation, i.e.\,within 0.3-0.5\,dex.
This scatter is smaller than the range spanned by our sample by a factor of $\approx$3, meaning that SED-dependent biases in our $M_\textrm{BH}$ cannot explain the observed trends presented in Section~\ref{sec:results:obs}.

\subsection{Quasar physics}

\subsubsection{AGN accretion models}
\label{sec:discuss_accretion}

In Section~\ref{sec:results:models}, we compared our observations with predictions from the \qsosed\ models of \citet{2018MNRAS.480.1247K}.
The predictions for $\alpha_\textrm{ox}$ made by these models have recently been tested over a much broader parameter space ($M_\textrm{BH}\approx 10^{7-10} M_\odot$ and $L_{3000} \approx 10^{43.5-47}\ergps$) by \citet{2022arXiv221011977M}, who find that the \qsosed\ model predicts the optical and X-ray SED fairly well for $M_\textrm{BH}\lesssim 10^{9} M_\odot$, but that at higher masses the outer accretion disc spectra are predicted to be too cool to match the observed data, especially at lower Eddington ratios. 
This finding is consistent with our result (in Fig.~\ref{fig:aox}) that the 2\,keV emission is predicted to be slightly stronger (relative to the 2500\,\AA\ emission) than observed at $M_\textrm{BH}\approx 10^{9.5} M_\odot$.

In this work we have also quantified the \ion{He}{II} emission, which provides a new, complementary probe of the ionizing SED across the $M_\textrm{BH}$--$L/L_\textrm{Edd}$ space. 
In Section~\ref{sec:results:models} we found that, for $L/L_\textrm{Edd} \gtrsim 0.1$, the observed \ion{He}{II} EW is qualitatively similar to the behaviour of the 54\,eV ionizing SED predicted by the \qsosed\ models.
The observations are consistent with a scenario in which (at least for $L/L_\textrm{Edd} \gtrsim 0.1$) the strength of \ion{He}{II} emission is set directly by the ionizing photon luminosity at 54\,eV, and thus that \ion{He}{II} is providing a probe of the EUV which is not directly observable.
Moreover, the observed \ion{He}{II} EW behaviour provides further evidence for the soft excess to be an intermediate, warm Comptonising component which behaves in the way in which the \qsosed\ models predict. The strongest and weakest 228\,\AA\ emission (relative to the 1640\,\AA\ continuum) are both produced at high Eddington ratios, at low ($\approx10^8M_\odot$) and high ($\approx10^{10}M_\odot$) SMBH mass respectively.

However, the match between the observed \ion{He}{II} and the predicted strength of the 54\,eV ionizing luminosity is not perfect, especially at $L/L_\textrm{Edd} \lesssim 0.1$. 
Intriguingly, this $L/L_\textrm{Edd}$ regime is similar to the region of the $M_\textrm{BH}$--$L_\textrm{2500\,\AA}$ space where \citet{2022arXiv221011977M} find a mismatch between the observed and predicted $\alpha_\textrm{ox}$,
which could suggest that the SED models are inaccurate in this part of the $M_\textrm{BH}$--$L/L_\textrm{Edd}$ parameter space.
However, this regime is also where the \ion{He}{II} and \civ\ EWs appear to be less well correlated, in the sense that the \civ\ EW increases towards the lowest Eddington ratios on our sample, while the \ion{He}{II} EW is observed to have more moderate values at $L/L_\textrm{Edd} < 0.1$.
The observations could therefore indicate
 a decoupling between the \ion{He}{II} EW and the 228\,\AA\ continuum flux at these Eddington ratios, perhaps if changes in the BLR covering factor lead to differences in the fraction of the continuum source which is reprocessed into emission lines. Another possibility is that the \ion{He}{II} continuum becomes optically thin, for instance if the density of the BLR were to decrease (which could indicate the absence of a dense outflow).
Either way, the observed switch in \ion{He}{II} behaviour above and below $L/L_\textrm{Edd} \approx 0.1$, which is not reflected in the \qsosed\ models, suggests that a fundamental change occurs in the structure of either the BLR or the accretion flow.

\civ\ is a resonant doublet with a more complicated ionic structure than \ion{He}{II}. However, the close correspondence of the \civ\ blueshift with the \ion{He}{II} EW, allied with the fact that the \ion{He}{II} behaviour can be consistently explained with trends in the SED,  suggests that the \civ\ morphology is governed by accretion physics - specifically the shape of the SED in the near and extreme ultraviolet regions.
Given the proximity of the relevant \civ\ and \ion{He}{II} ionization edges,
at 48 and 54\,eV respectively, this result is perhaps unsurprising.

\subsubsection{AGN outflow models}
\label{sec:discuss_outflows}

In this subsection we test the predictions made by \citet{2019A&A...630A..94G}, who summarize current understanding of AGN accretion and outflow mechanisms with a particular focus on the physical conditions required to drive powerful winds from the accretion disc through radiation line driving.
We note again that the picture described by \citet{2019A&A...630A..94G} might not be the only plausible model for AGN outflows, but we choose to compare with their picture as it provides clear testable predictions within a well-defined framework.
In particular, \citet{2019A&A...630A..94G} suggest that both $L/L_\textrm{Edd} \gtrsim 0.25$ and $M_\textrm{BH} \gtrsim 10^8 M_\odot$ are required to power strong outflows from AGN through radiation line driving:  
below these thresholds the X-ray flux is strong enough to over-ionize material and the ultraviolet flux will be too low to accelerate a line-driven wind.

For the purposes of this comparison, we assume that any blue-wing excess in the \civ\ emission line profile is tracing an outflow along the line-of-sight from the accretion disc, and hence that the blueshift presented in Fig.~\ref{fig:CIV} is a measure of the strength of emission from the outflowing wind \citep{2004ApJ...611..107L, 2011AJ....141..167R}. 
The origin of the \civ\ emission line blueshift is still debated \citep[see][for an alternative view]{2013ApJ...769...30G, 2016Ap&SS.361...67G}, but a growing body of work is connecting the \civ\ emission morphology with more unambiguous tracers of line driven winds. For example, the strengths and velocities of broad \civ\ absorption troughs have been shown to correlate with the \civ\ emission blueshift \citep{2020MNRAS.492.4553R, 2022ApJ...939L..24R}, and the velocities of narrow \civ\ line-locked `triplet' absorption features are also strongly correlated with the emission blueshift (Rankine et al. in preparation).

For the discussion in this subsection, we therefore assume that objects with larger \civ\ blueshifts have stronger disc winds.
To be more precise, the \civ\ blueshift is taken as a measure of the strength of emission from outflowing gas relative to the strength of emission from virialized gas at the systemic redshift.
In this paradigm, the  trends seen in Fig.~\ref{fig:CIV} are in good agreement with the picture proposed by \citet{2019A&A...630A..94G}. We see large ($\gtrsim1000$\,\kmps) median \civ\ blueshifts  only in bins with  high SMBH masses \textit{and}  high Eddington ratios. Furthermore, we do indeed see a more complicated mass dependence above $L/L_\textrm{Edd} \approx 0.2$.
High $L/L_\textrm{Edd}$ is therefore a necessary, but not sufficient, condition for observing large \civ\ blueshifts, consistent with the results of \citet{2005MNRAS.356.1029B}.

In detail, we only observe strong outflow signatures in objects with $M_\textrm{BH} \gtrsim 10^9 M_\odot$, which is somewhat higher than the criterion of $M_\textrm{BH} \gtrsim 10^8 M_\odot$ proposed by \citet{2019A&A...630A..94G}.
Requiring $M_\textrm{BH} \gtrsim 10^9 M_\odot$ and $L/L_\textrm{Edd} \gtrsim 0.2$ together ensures that the criterion $L_\textrm{bol} \gtrsim 10^{45.5}\ergps$ is satisfied. Above this $L_\textrm{bol}$ threshold, \citet{2014MNRAS.442..784Z} suggest that quasar winds are capable of driving ionized gas (as traced by [\ion{O}{III}]\,$\lambda$5008 emission) beyond the escape velocity of the host galaxy. The kinematics of \civ\ and [\ion{O}{III}] are known to correlate \citep{2019MNRAS.486.5335C}, and our observed \civ\ blueshift behaviour is therefore consistent with the conclusion of \citet{2014MNRAS.442..784Z} that $L_\textrm{bol} \gtrsim 10^{45.5}\ergps$ is required for quasar feedback to operate.

For $10^8 M_\odot \lesssim M_\textrm{BH} \lesssim 10^9 M_\odot$ and $L/L_\textrm{Edd} \gtrsim 0.1$, we see the strongest \ion{He}{II} and strongest non-outflowing \civ\ line emission. One possible explanation for this behaviour would be that this emission represents ionized material which has been launched from the accretion disc, but lacks the ultraviolet luminosity to accelerate the outflow, meaning that such material falls back and virializes instead of escaping.
In such a scenario the strong symmetric emission from high-ionization ultraviolet lines would represent a failed line-driven wind, analogous to models of the low-ionization BLR which represent a failed dust-driven wind (\citealt{2011A&A...525L...8C, 2018MNRAS.474.1970B}, see also \citealt{2017ApJ...847...56E}).

While we observe a reasonably good qualitative agreement between the \civ\ blueshift behaviour and the \citet{2019A&A...630A..94G} predictions for line-driven winds, the reality is likely more complicated.
In particular, \citet{2019A&A...630A..94G} do not consider any emission from a `soft excess'. Instead they assume that the ionizing SED consists of just two components, emitted from a thermal disc and a hot corona. 
Such a simple model is unlikely to explain our observational results: the different behaviour of \ion{He}{II} EW and $\alpha_\textrm{ox}$ as a function of $M_\textrm{BH}$ and $L/L_\textrm{Edd}$ points to the presence of a third spectral component in the EUV which can vary separately from the disc and corona.

Other physical effects could also be at play. In particular, as the accretion rate increases above $\dot{m} \gtrsim 0.3$, we expect the disc structure to transition between geometries akin to slim discs and thin discs \citep{1988ApJ...332..646A, 2013LRR....16....1A}.
At low accretion rates, slim discs are well approximated by the \citet{1973blho.conf..343N} thin disc solution, as used in \qsosed, but we expect this to be less accurate as $\dot{m}$ increases.
In other words, the regime in which \qsosed\ appears to best match our data is also the regime in which we might expect it to be least accurate.
The origin of the apparent transition around $L/L_\textrm{Edd} \approx 0.1$ in Figs.~\ref{fig:CIV} and \ref{fig:HeII} is therefore still uncertain and further work is required to fully understand the interplay between AGN accretion flows, the ionizing SEDs they produce, and the outflows they drive.

Line-driven winds from high Eddington ratio AGN are often cited as a potentially important component of radiative-mode (quasar-mode) feedback \citep{2012ApJ...745L..34Z}. While difficult to observe directly, such feedback modes are required to regulate galaxy growth and explain the tight SMBH-galaxy correlations observed in the local universe \citep[see][for a review]{2012ARA&A..50..455F}.
However, most SMBHs in the local universe do not have masses above $10^9M_\odot$,  so our results might suggest that the line-driven winds traced by \civ\ cannot have a significant effect on their host galaxies' growth and co-evolution as they never reach the SMBH masses required to launch strong winds.
There are at least two solutions to this apparent problem. First is that radiative-mode feedback could still be operating through ionized gas outflows, but that the gas is in a different ionization phase and is not seen in \civ, but instead in other bands such as the X-ray  `ultra-fast outflows' \citep[][]{2021NatAs...5...13L}. Second could be that quasar-mode feedback is only effective when coupled to dusty gas \citep{2008MNRAS.385L..43F, 2018MNRAS.479.3335I, 2022ApJ...938...67R}, thus having most impact when the AGN is obscured by dust \citep[][]{2019MNRAS.487.2594T, 2020MNRAS.495.2652L, 2021ApJ...906...21J, 2022ApJ...934..101A}.

\subsubsection{SMBH spin and system inclination}
\label{sec:spin_inc}
In our \qsosed\ modeling, we kept inclination fixed at $\cos i=0.5$ and only presented the non-spinning SMBH case, $a_*=0$. However, both of these parameters have an impact on the predicted SEDs. The impact of SMBH spin is particularly pronounced; plots matching those in the right-hand panels of Figs.~\ref{fig:aox} and \ref{fig:HeII} are presented in Appendix~\ref{appendixc}. The basic finding from the maximal spin models is that the observed trends of $\alpha_\textrm{ox}$ with mass and Eddington fraction are not reproduced, for reasons that are explained in Appendix~\ref{appendixc}. In fact, all of the maximal spin models have $\alpha_\textrm{ox} \gtrsim -1.5$, meaning that the X-ray luminosity is always quite high compared to the optical and ultraviolet, and the observed soft spectra at high mass and high Eddington fraction are not reproduced for $a_*=1$. If there are a significant proportion of maximally spinning SMBHs in our quasar sample, this would imply that the model predictions are not valid for high spin objects, potentially undermining many of the results discussed in Section~\ref{sec:discuss_accretion}. Alternatively, if the \qsosed\ models are correct, the good agreement at low spin and poor agreement at high spin would imply that most SDSS quasars at $z\approx2$ typically have low or moderate SMBH spins.

SMBH spin is most commonly estimated from broad iron line emission in the X-ray band \citep{reynolds2019}. Spin measurements tend to be rather high, with the majority of X-ray measurements in AGN consistent with maximally spinning SMBHs. This apparent preference might initially appear to be inconsistent with our results. However, there are a number of factors at work. First, discs around maximally spinning SMBHs have higher radiative efficiencies and are thus more luminous. As shown in figure~3 of \cite{reynolds2019}, this might lead to high spins being over-represented in a sample. One could also imagine further selection effects if spins are easier to measure when they are close to maximal and the iron line is broader. Second, the majority of spin measurements are at lower masses ($M_\textrm{BH}\lesssim 10^{8}~M_\odot$) than in our sample, with only a handful of spin measurements in our considered mass regime. In fact, there is some tentative evidence for a decrease of SMBH spin with increasing mass \citep{siskreynes2022}, behaviour that is expected if accretion is coherent at low masses and more incoherent at higher masses, as predicted by both semi-analytic models and hydrodynamic simulations \citep{king2008,sesana2014,zhang2019,bustamante2019}. If our results do indeed favour low or moderate SMBH spins in luminous quasars, then they are consistent with this proposed trend, especially since it is the high-mass quasars which cannot be matched by the maximally spinning \qsosed\ models. Any conclusions drawn here should be treated with caution, given (i) the difficulties in obtaining reliable and unbiased spin measurements, and (ii) the lack of knowledge about the impact of spin on the EUV and X-ray regions of the quasar SED. 

We fixed inclination in the \qsosed\ models, adopting $\cos i=0.5$. Following Copernican reasoning, we expect AGN to have a random, isotropic distribution of viewing angles, in which case the mean viewing angle of all AGN is $\cos i=0.5$. Factoring in obscuration by a putative `torus' beyond some maximum inclination, and foreshortening/limb darkening of the disc continuum might be expected to bias this mean viewing angle to lower (more face-on) inclinations \citep[e.g.][]{krolik1998,matthews2017}.
%In future, one could consider a Monte Carlo approach in which a population of simulated quasars is produced for a distribution of viewing angles, and the median SED properties are plotted in the right-hand panels of Figs.~\ref{fig:aox} and \ref{fig:HeII}. A similar route could be taken for SMBH spin assuming some probability distribution function for $a_*$. However, both these approaches would introduce additional free parameters, and would be at odds with the approximate spirit of our comparison between models and observational data. We can, however, discuss the sensitivity of the model SEDs to inclination, to give a feel for how the results would change if, say, a lower inclination were adopted. 
If we were to adopt a different inclination in \qsosed, we can think about how the model predictions would change.

The hard X-ray source in \qsosed\ is isotropic, whereas the warm and thermal components have a disc-like geometry and thus produced an observed luminosity $\propto \cos i$, such that lower inclinations have higher luminosities. The impact of inclination on the outputs from \qsosed\ can thus be straightforwardly understood. Changing inclination from $i_1 \to i_2$ results in a fractional change in $L_{3000}$ of $(\cos i_2 / \cos i_1)$. Since, for a given input $\dot{m}$, we calculate $L/L_{\rm Edd}$ from $L_{3000}$, we obtain a linear scaling of the $y$-axis of the right-hand panels of Figs~\ref{fig:HeII} and \ref{fig:aox} by the same factor. The change in $\alpha_{\rm ox}$ is $\Delta \alpha_{\rm ox} = -0.3838 \log_{10} (\cos i_2 / \cos i_1)$; adopting a lower inclination with $\cos i =0.75$ would result in a more negative $\alpha_{\rm ox}$ in all simulation bins by $\approx 0.07$. Finally, the ratio $L_{228}/L_{1640}$ undergoes small changes with $\cos i$, but these are fairly uniform across the simulation grid and thus unimportant, given that the proportionality constant between $L_{228}/L_{1640}$ and He~\textsc{ii} EW is not known. Furthermore, the He~\textsc{ii} EW depends on the $L_{228}$ seen by the He~\textsc{ii} gas, rather than the $L_{228}$ seen by a distant observer, meaning that the true inclination dependence of He~\textsc{ii} EW would depend on the BLR geometry. 
We have explicitly checked that the anticipated changes in $L_{3000}$ and $\alpha_{\rm ox}$ are indeed reproduced in \qsosed, except for small departures in the $\alpha_{\rm ox}$ due to contamination of the $2~{\rm keV}$ flux by the warm component.

\subsection{Wider implications and future work}
\label{sec:discuss:future}

\subsubsection{Comparison with other populations}

Recent work has attempted to compare the ultraviolet emission properties in high redshift ($z\gtrsim6$) quasars with their lower redshift analogues \citep{2017ApJ...849...91M, 2019MNRAS.487.3305M, 2019ApJ...873...35S, 2020ApJ...905...51S,  2021ApJ...923..262Y, 2022MNRAS.513.1801L, 2022ApJ...925..121W}. Such quasars are (by selection) very luminous, and generally display large \civ\ blueshifts. 
From our results, we would argue that comparative studies should match AGN samples not just in luminosity, but in two independent parameters which trace $L/L_\textrm{Edd}$ and $M_\textrm{BH}$. 
Stepney et al. (in preparation) will discuss this further in a study of \civ\ and \ion{He}{II} emission in SDSS quasars with redshifts $z>3.5$.

Current samples of $z\gtrsim6$ quasars include a significant number of objects with inferred $L/L_\textrm{Edd}>1$, which lie outside the parameter space explored in this work. We have verified that the sample of 37 quasar spectra with redshifts $z>6.3$ presented by \citet{2021ApJ...923..262Y} typically show narrower \mgii\ profiles than $z\approx2$ SDSS objects with the same 3000\,\AA\ luminosities, suggesting smaller $M_\textrm{BH}$ and larger $L/L_\textrm{Edd}$ (for a given $L_{3000}$) than the quasars characterized in this work. For such objects it is therefore not surprising that their typical ultraviolet emission properties are different from the majority of the SDSS population at $z\approx2$.

The X-ray properties of the $z>6$ quasar population appear to be consistent with the $\alpha_\textrm{ox}$--$L_\textrm{2500\,\AA}$ relation seen at lower redshifts \citep{2017A&A...603A.128N, 2019A&A...630A.118V, 2020MNRAS.491.3884P, 2021MNRAS.501.6208P}.
To gain further insight, we encourage future works to consider the behaviour of $\alpha_\textrm{ox}$ as a function of both $M_\textrm{BH}$ and $L/L_\textrm{Edd}$ simultaneously.

\subsubsection{BLR metallicities}

The relative strengths of many ultraviolet emission lines are known to correlate with the \civ\ and \ion{He}{II} emission properties \citep{2011AJ....141..167R, 2020MNRAS.496.2565T}.
In particular, the flux ratios of high ionization ultraviolet lines such as \civ, \ion{N}{V}\,$\lambda$1240 and \ion{Si}{IV}\,$\lambda1400$ are tightly correlated with the \civ\ blueshift and \ion{He}{II} EW \citep{2021MNRAS.505.3247T}.
Assuming no changes in density or ionization structure or geometry of the BLR, changes in these line ratios are sometimes taken to reflect changes in the metal content of the BLR
\citep{2006A&A...447..157N}.
Such an interpretation, combined with the results in this work, gives rise to a paradigm where the metal content of quasar BLRs is largest in objects with the largest $M_\textrm{BH}$ and $L/L_\textrm{Edd}$, as noted by \citet{2018MNRAS.480..345X} and \citet{2021ApJ...910..115S} respectively.

However, as shown in \citet{2021MNRAS.505.3247T}, the variation in these line ratios can be explained with changes in the density of the emitting gas, and need not involve changes in metallicity \citep[see also appendix A4 of][]{2006ApJ...637..157C}.
In particular, the ultraviolet emission line ratios seen in objects with large \civ\ blueshifts can be explained by emission from relatively dense gas which is located closer to the ionizing source, while the line ratios in objects with high EW, symmetric \civ\ emission are consistent with emission from less dense gas at larger radii.
Given the trends seen in Figs.~\ref{fig:CIV} and \ref{fig:HeII}, this is a much more natural explanation:
objects with different SMBH masses and accretion rates have different accretion flows, which give rise to different EUV SEDs (as traced by \ion{He}{II}) and different kinematic and density structures in the BLR (traced by the \civ\ blueshift and high ionization line ratios respectively).
Under this alternative paradigm the BLR metallicity would be free to vary independently of $M_\textrm{BH}$ and $L/L_\textrm{Edd}$, and need not be super-solar in the early universe \citep[cf.][]{2022MNRAS.513.1801L, 2022ApJ...925..121W}.

\subsubsection{Quasar cosmology}

Quasars are visible out to large cosmological distances, and display remarkably homogeneous behaviour across cosmic time. 
A growing body of work has proposed the non-linear scaling between the ultraviolet and X-ray continuum fluxes
(i.e.\ the $\alpha_\textrm{ox}$--$L_\textrm{2500\,\AA}$ relation) as a way to use quasars as standardizable candles for cosmological measurements 
\citep{2015ApJ...815...33R, 2017AN....338..329R, 2019NatAs...3..272R, 2017A&A...602A..79L, 2019A&A...631A.120S, 2020A&A...642A.150L, 2022A&A...663L...7S, 2022MNRAS.510.2753K}.
However, \citet{2022ApJ...935L..19P} have recently raised concerns about such methods. In this work we have shown that the ultraviolet emission lines provide further information which could be used to mitigate such concerns.
With knowledge of the \ion{Mg}{II} velocity width, and either the \ion{He}{II} strength or the \ion{C}{IV} properties, one should be able to locate an object in the $M_\textrm{BH}$--$L/L_\textrm{Edd}$ plane, and hence infer the intrinsic luminosity in a cosmology-independent way. By comparing to the observed fluxes one could then (in principle) infer a constraint on the Hubble parameter $H(z)$.
However, further work is still required. In particular, we need to build a sample of quasars with ultraviolet emission line measurements which have independent measurements of the luminosity distance, in order to calibrate our $M_\textrm{BH}$--$L/L_\textrm{Edd}$ space in a cosmology-independent manner, in an analogous way to the use of the `inverse distance ladder' to calibrate type Ia supernovae as standard candles \citep[e.g.][]{2001ApJ...553...47F, 2021ApJ...908L...6R}.

\subsubsection{Time variability and upcoming surveys}

Changes in $L/L_\textrm{Edd}$ for a quasar with fixed $M_\textrm{BH}$ will lead to changes in the emitted spectrum, but such changes in SMBH fueling are expected to generally occur on the viscous time-scale, which is on the of order of tens to thousands of years.
However, SMBH accretion is inherently stochastic and the emitted flux varies by a factor of a few on shorter time-scales of just years. The time-scale and amplitude of this  intrinsic `flickering' are now known to depend on the SMBH mass and accretion rate \citep{2022ApJ...936..132Y}, and this stochastic flickering will contribute to the scatter within each binned region of our parameter space (Section~\ref{sec:results:obs}).

In terms of spectroscopic variability, \citet{2020ApJ...899...96R} showed that individual SDSS-RM quasars with multiple epochs of spectroscopy (i.e. with fixed $M_\textrm{BH}$) can vary in essentially every direction in the \ion{C}{IV} blueshift--EW space, although objects with large blueshifts tend to show a change in blueshift and objects with strong EW show a change in EW.
In the near future, SDSS-V \citep{2017arXiv171103234K} will provide multi-epoch spectroscopic data for tens of thousands of luminous quasars, providing new insights into AGN variability.

At the same time, surveys such as DESI \citep{2022arXiv220808517A, 2022arXiv220808511C} and 4MOST \citep{2019Msngr.175...42M, 2022arXiv221107324E}
will probe fainter, yielding spectra of lower luminosity quasars than the sample investigated in this work, and future data releases from the \textit{eROSITA} all-sky survey will include X-ray flux measurements for millions of AGN. Together these surveys will provide new constraints on the spectroscopic properties and ionizing SEDs of luminous AGN across the $M_\textrm{BH}$--$L/L_\textrm{Edd}$ parameter space.

\section{Conclusions}
\label{sec:conclude}
%The last numbered section should briefly summarise what has been done, and describe
%the final conclusions which the authors draw from their work.

We have investigated the rest-frame ultraviolet emission line properties in 186\,303 SDSS quasars with redshifts $1.5<z<2.65$.
We can infer $\alpha_\textrm{ox}$, the logarithmic ratio of the rest-frame 2\,keV and 2500\,\AA\ luminosities, for 5031 quasars in our sample.
Using the FWHM of \mgii\,$\lambda$2800 as a proxy for the virial velocity, we quantify the average properties of the \civ\,$\lambda1549$ and \ion{He}{II}\,$\lambda$1640 emission lines across the two-dimensional space spanned by $M_\textrm{BH}$ and $L/L_\textrm{Edd}$, and use these observations to confront qualitative predictions of when radiation-driven outflows should dominate kinetic feedback mechanisms \citep{2019A&A...630A..94G} and theoretical SEDs arising from models of AGN accretion flows \citep{2018MNRAS.480.1247K}. Our main conclusions are:
\begin{enumerate}
    \item As shown in previous works \citep{2011AJ....141..167R, 2020MNRAS.492.4553R}, the blueshift and EW of \civ\ correlate with the EW of \ion{He}{II}. Objects with strong \ion{He}{II} have high EW \civ\ with little or no blue excess, while objects with weaker \ion{He}{II} show smaller EW \civ\ with larger \civ\ blueshifts.
    
    \item We recover a Baldwin effect, but instead of simply correlating with the ultraviolet luminosity, we find that the \civ\ and \ion{He}{II} properties display more complicated trends in the $M_\textrm{BH}$--$L/L_\textrm{Edd}$ plane.   The dynamic range in \ion{He}{II} EW is greatest at Eddington ratios  $\gtrsim$0.1 (Fig.~\ref{fig:HeII}). The largest \civ\ blueshifts are only observed at  high $L/L_\textrm{Edd}$ \textit{and} high $M_\textrm{BH}$, while the highest  EWs are seen only at high $L/L_\textrm{Edd}$ and relatively low $M_\textrm{BH}$ (Fig.~\ref{fig:CIV}). Composite spectra from these two extrema are shown in blue and green in Fig.~\ref{fig:sample}.
    
    \item In contrast to the ultraviolet emission line properties, but consistent with previous work in the literature, $\alpha_\textrm{ox}$ displays a simpler behaviour across the $M_\textrm{BH}$--$L/L_\textrm{Edd}$ plane (Fig.~\ref{fig:aox}), albeit in a much smaller sample. $\alpha_\textrm{ox}$ shows a more direct correlation with the ultraviolet continuum luminosity than the emission lines,
    although $\alpha_\textrm{ox}$ does show some dependence on $M_\textrm{BH}$ at fixed $L_{3000}$.
    Future data releases from eROSITA, SDSS-V and 4MOST will increase the number of known quasars with X-ray data.
    
    \item Under the assumption that blueshifted \civ\ emission is tracing a disc wind accelerated by radiation line driving, we find our results are consistent with the global scheme for accretion and outflow mechanisms proposed by \citet{2019A&A...630A..94G}. In particular, an Eddington-scaled mass accretion rate  $\dot{m}\gtrsim0.25$ is required for the formation of the strongest line-driven winds. \citet{2019A&A...630A..94G} suggest that $M_\textrm{BH} > 10^8M_\odot$ is also required to launch strong line-driven winds, however we only observe the largest \civ\ blueshifts in objects with \mgii-inferred $M_\textrm{BH} \gtrsim 10^9M_\odot$.
    Strong line emission at $M_\textrm{BH} \lesssim 10^9M_\odot$ could perhaps indicate a `failed' line-driven wind.
    
    \item Absent large changes in the density or geometry of the broad line region, the strength of \ion{He}{II} is probing the strength of 54\,eV ionizing radiation in the `unseen' portion of the ultraviolet SED. Above $L/L_\textrm{Edd}\approx0.1$, we find that the EW of \ion{He}{II} is broadly consistent with the \citet{2018MNRAS.480.1247K} \qsosed\ model. In other words, the relative strength of the 54\,eV flux (which is photoionizing the broad line region) compared to the 1640\,\AA\ continuum is consistent with a relatively simple model where 
    the strength of the `soft excess' is adjusted to give the correct bolometric luminosity while keeping the strength of the hot coronal emission fixed at two per cent of the Eddington luminosity  \citep[as proposed by][]{2018MNRAS.480.1247K}.
        
    \item Below $L/L_\textrm{Edd}\approx0.1$, something changes in the physics of the broad line region, with no strong \civ\ blueshifts observed and much weaker trends in \ion{He}{II}. The \qsosed\ models do not provide as good a match to the observed \ion{He}{II} EWs, consistent with the results of \citet{2022arXiv221011977M} who find a discrepancy between the observed and predicted  $\alpha_\textrm{ox}$ in the same region of  $M_\textrm{BH}$--$L/L_\textrm{Edd}$ parameter space. 
    
    \item Similar to \citet{2022arXiv221011977M}, we also find no strong evidence for high SMBH spins in our quasar sample: the zero-spin  \qsosed\ models provide an acceptable match to the SED probes across a significant portion of our observed parameter space while the maximally spinning models do not.
    If a significant fraction of our quasar sample have maximally spinning SMBHs, this would suggest that the \qsosed\ model assumptions are not valid for high spin objects. Alternatively, taking the model results at face value would suggest low or moderate spins in typical SDSS quasars at $z\approx2$. 
\end{enumerate}

\section*{Acknowledgements}
%The Acknowledgements section is not numbered. Here you can thank helpful
%colleagues, acknowledge funding agencies, telescopes and facilities used etc.

We gratefully acknowledge useful discussions with Chris Done and Jake Mitchell. MJT thanks  Chiara Mazzucchelli,  Claudio Ricci and  Roberto Assef for insightful comments, and Jinyi Yang for sharing the sample of $z>6$ quasar spectra from \citet{2021ApJ...923..262Y}.
We thank  Margherita Giustini, Ari Laor, Elisabeta Lusso and an anonymous referee for their useful feedback on the submitted manuscript.

MJT acknowledges support from a FONDECYT postdoctoral fellowship (3220516). JHM acknowledges funding from the Royal Society. ALR acknowledges support from UKRI (MR/T020989/1).
This work  made use of \textsc{astropy} \citep{astropy:2013, 2018AJ....156..123A, 2022ApJ...935..167A},  \textsc{matplotlib} \citep{Hunter:2007}, and \textsc{numpy} \citep{numpy}.

Funding for the Sloan Digital Sky Survey IV has been provided by the Alfred P. Sloan Foundation, the U.S. Department of Energy Office of Science, and the Participating Institutions. SDSS-IV acknowledges
support and resources from the Center for High-Performance Computing at
the University of Utah. The SDSS web site is www.sdss.org.

SDSS-IV is managed by the Astrophysical Research Consortium for the 
Participating Institutions of the SDSS Collaboration including the 
Brazilian Participation Group, the Carnegie Institution for Science, 
Carnegie Mellon University, the Chilean Participation Group, the French Participation Group, Harvard-Smithsonian Center for Astrophysics, 
Instituto de Astrof\'isica de Canarias, The Johns Hopkins University, Kavli Institute for the Physics and Mathematics of the Universe (IPMU) / 
University of Tokyo, the Korean Participation Group, Lawrence Berkeley National Laboratory, 
Leibniz Institut f\"ur Astrophysik Potsdam (AIP),  
Max-Planck-Institut f\"ur Astronomie (MPIA Heidelberg), 
Max-Planck-Institut f\"ur Astrophysik (MPA Garching), 
Max-Planck-Institut f\"ur Extraterrestrische Physik (MPE), 
National Astronomical Observatories of China, New Mexico State University, 
New York University, University of Notre Dame, 
Observat\'ario Nacional / MCTI, The Ohio State University, 
Pennsylvania State University, Shanghai Astronomical Observatory, 
United Kingdom Participation Group,
Universidad Nacional Aut\'onoma de M\'exico, University of Arizona, 
University of Colorado Boulder, University of Oxford, University of Portsmouth, 
University of Utah, University of Virginia, University of Washington, University of Wisconsin, 
Vanderbilt University, and Yale University.

For the purpose of open access, the authors will apply a Creative Commons Attribution (CC BY) licence to any Author Accepted Manuscript version arising from this submission.

%%%%%%%%%%%%%%%%%%%%%%%%%%%%%%%%%%%%%%%%%%%%%%%%%%
\section*{Data Availability}
%The inclusion of a Data Availability Statement is a requirement for articles published in MNRAS. 

The spectroscopic data underlying this article are available from SDSS.\footnote{\url{https://www.sdss4.org/dr17/}} The X-ray data sets underlying this article are available from the references given in Section~\ref{sec:Xray_data}. The composite spectra in Fig.~{\ref{fig:sample}} are available as online supplementary material.



\bibliographystyle{mnras}
\bibliography{refs.bib}

%%%%%%%%%%%%%%%%% APPENDICES %%%%%%%%%%%%%%%%%%%%%
\appendix

% Temple_Matthews_etal_22/xray_sed_bis/SEDs.vsz
\section{X-rays and BLR photoionization}
\label{appendixa}

\begin{figure}
	\includegraphics[clip=true, trim={0 30 0 10}, width=\columnwidth]{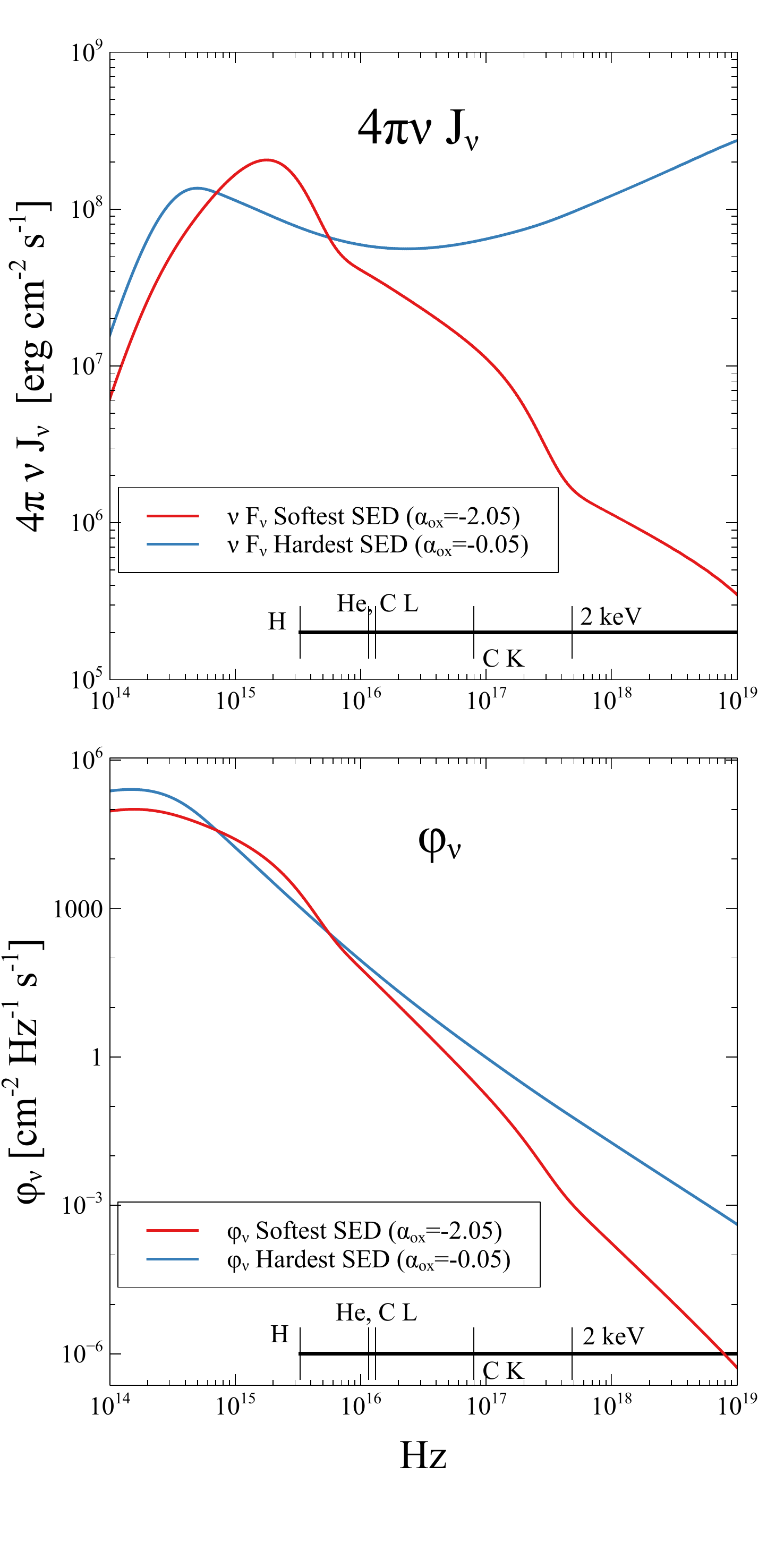}
    \caption{The upper panel shows the hardest and softest SEDs presented
    in this paper. The lower panel shows the flux of photons $\phi_\nu$ for the same models.
    Some important energies are indicated by the hashed lines near the bottom of each panel. 
    These show the ionization potentials of ground-state H$^0$ and
    He$^{2+}$, the L and K shells of C$^{2+}$, and  2\,keV.
    The flux of ionizing photons is orders of magnitude smaller at 2\,keV compared to the  flux at the \ion{He}{II} and \civ\ photoionization edges
    even for the hardest SED.
    }
    \label{fig:SEDphi}
\end{figure}

This Appendix shows that the X-ray portion of the
SED is an insignificant source of ionization 
for typical AGN emission line regions. 
This is surprising since the SEDs shown in Fig.~\ref{fig:qso_seds} 
have a significant fraction of their power at high energies, and the
 ultraviolet--X-ray hardness ratio $\alpha_\textrm{ox}$ is known to correlate with emission line properties.
However, photoionization is photon-counting,
and there are relatively few X-ray photons despite their significant energy.

To illustrate this point, we use the hardest and softest SEDs from our model grid (Section~\ref{sec:models}), corresponding to $\dot{m}=0.027$ and $\dot{m}=1.000$ at $M_\textrm{BH}=10^{10}M_\odot$.
These SEDs are shown in the top panel of Figure \ref{fig:SEDphi}, and have $\alpha_\textrm{ox}=-0.05$ and
$\alpha_\textrm{ox}=-2.05$ respectively.

The photoionization rate for a given shell $n$ is
\begin{equation}
 \Gamma_n = \int_{\nu_0}^{\infty} \sigma_{\nu} \phi_{\nu}\ d\nu  [{ \rm s}^{-1}]
 \label{eq:PhotoRate}
\end{equation}
where $\nu_0$,  $\sigma_{\nu}$, and $\phi_{\nu}$ are the photoionization threshold of 
shell $n$, its energy-dependent photoionization cross section  [cm$^{2}$] ,
and the flux of ionizing photons [cm$^{-2}$ s$^{-1}$ Hz$^{-1}$] \citep{2006agna.book.....O}.  
The total photoionization rate is the sum over all shells,
\begin{equation}
 \Gamma_{\rm total} = \sum_n \Gamma_n  [{ \rm s}^{-1}]
 \end{equation}

The flux of ionizing photons $\phi_{\nu}$ enters in the photoionization rate (Eq.~\ref{eq:PhotoRate}).  
This is the ratio $\phi_{\nu} = 4 \pi \nu J_{\nu} / (h\nu^2) $
and is shown in the lower panel of Fig.~\ref{fig:SEDphi}.  
The photon flux near 2\,keV  is typically 
$\sim$6\,dex fainter than the value near the peak. 

\begin{figure}
	\includegraphics[width=\columnwidth]{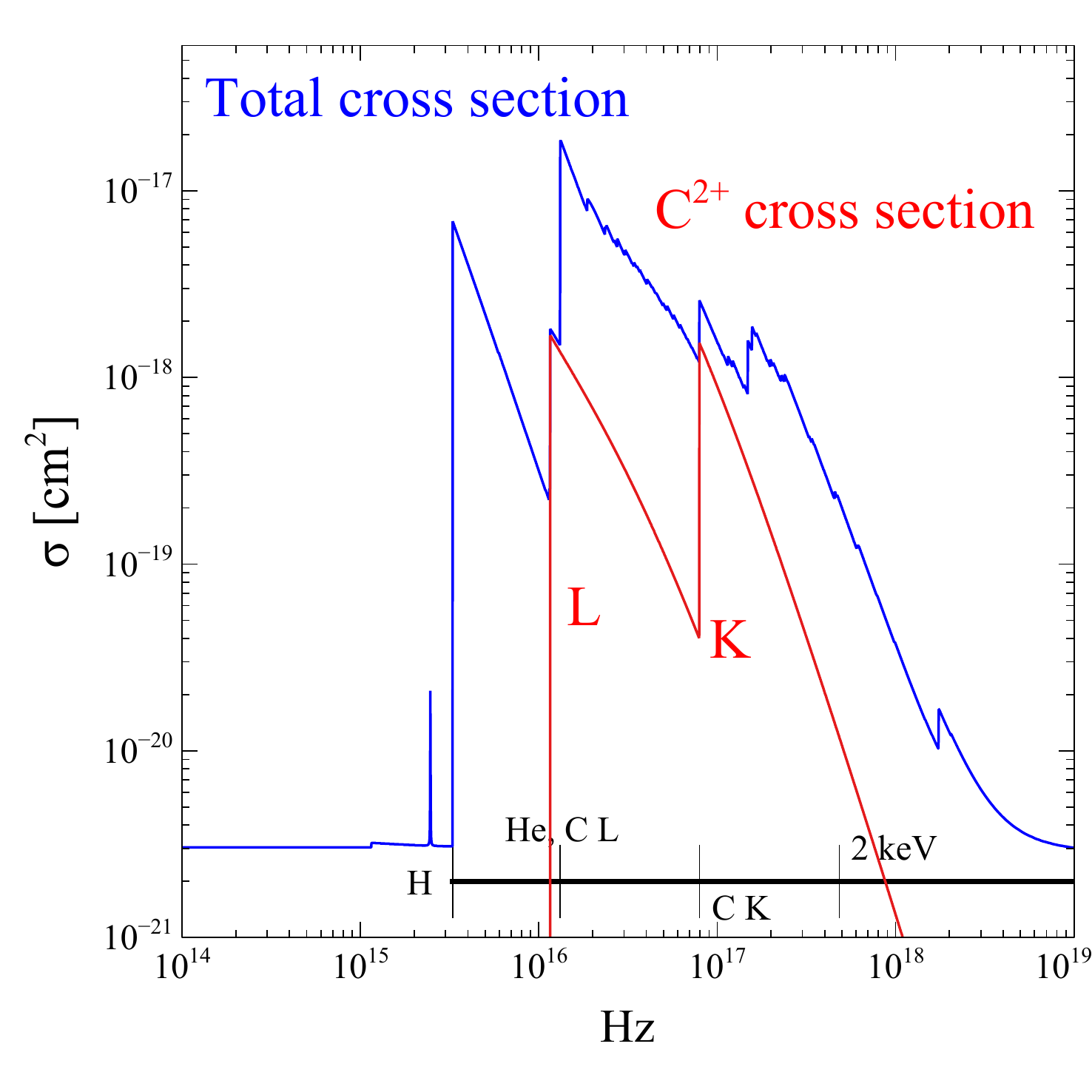}
    \caption{The K and L shell photoionization cross sections
    for C$^{2+}$ are shown as a function of energy
    as the red line.
    The cross section are from the calculations by
    \citet{1996ApJ...465..487V}.
    The blue line shows the total (absorption and scattering) gas opacity
    rescaled to match the C$^{2+}$ cross sections.
    The absorption opacity and the flux of photons
    (lower panel of Figure \ref{fig:SEDphi}),
    determine the photoionization rate
    (Equation \ref{eq:PhotoRate})
    and the effects of light upon matter
    shown in Figure \ref{fig:PhiSigma}.
    }
    \label{fig:photocs}
\end{figure}

The photon flux is multiplied by the photoionization
cross section to derive the photoionization rate
(Eq.~\ref{eq:PhotoRate}).
We  concentrate on C$^{2+}$ since photoionization
of that ion produces C$^{3+}$ and \civ\,$\lambda$1549 emission.
The shell-dependent cross sections for photoionization
of C$^{2+}$, taken from
\citet{1996ApJ...465..487V},
are shown in Fig.~\ref{fig:photocs}.
Both the 1s$^2$ K shell in the X-ray and the lower energy
2s$^2$ L shell are shown.
Both shells have two electrons and, as expected,
the peak photoionization cross sections are similar.
 
 %workarea Temple_Matthews_etal_22/xray_sed/photocs.vsz
 % this path is correct, not in xray_sed_bis

The photoionization rates for H$^0$ and
the two shells of C$^{2+}$ are listed in 
Table~\ref{tab:SEDgammas}.
The C$^{2+}$ K-shell rate is 17 to 60 times
smaller then the L-shell rate. 
Both are $\sim$10-30 times smaller than
the H$^0$ photoionization rate.
From this comparison we expect the 
ultraviolet continuum to play a more important
role than the X-rays in the photoionization of the BLR,
mainly due to the larger number of
softer photons.

\begin{table}
 \centering
 \caption{\label{tab:SEDgammas}  Photoionization rates for
 H$^0$ and the K and L shells of C$^{2+}$
 the hardest and softest SEDs.
 This allows us to compare the K-shell 
 photoionization rate, produced by X-rays, with the
 L-shell rates, the result of the ultraviolet part of the SED.
 The photoionization rates are given for both the hardest and softest SED considered here.
 For both SEDs, the L-shell rates are more than
 1\,dex larger than K-shell rates, showing
 that the X-ray portion of the SED has comparatively little effect on the
 photoionization of C$^{2+}$.}
 %these change font size making smaller than default
 % scope is only in table so does not need to be redone
 \null\smallskip
 \footnotesize
 \renewcommand\arraystretch{0.65}
 \begin{tabular}{ c c c c }
 \hline\hline
 Shell & $\Gamma_{\rm Softest}$ [s$^{-1}$] &  $\Gamma_{\rm Hardest}$ [s$^{-1}$] \\
 \hline
H$^0$ K & 8.24e+00 & 5.52e+00 \\
C$^{2+}$ L &2.66e-01 & 4.66e-01 \\
C$^{2+}$ K & 4.49e-03 & 2.71e-02\\
 \hline
 \end{tabular}
 \end{table}

\begin{figure}
	\includegraphics[clip=true, trim={0 0 0 10}, width=\columnwidth]{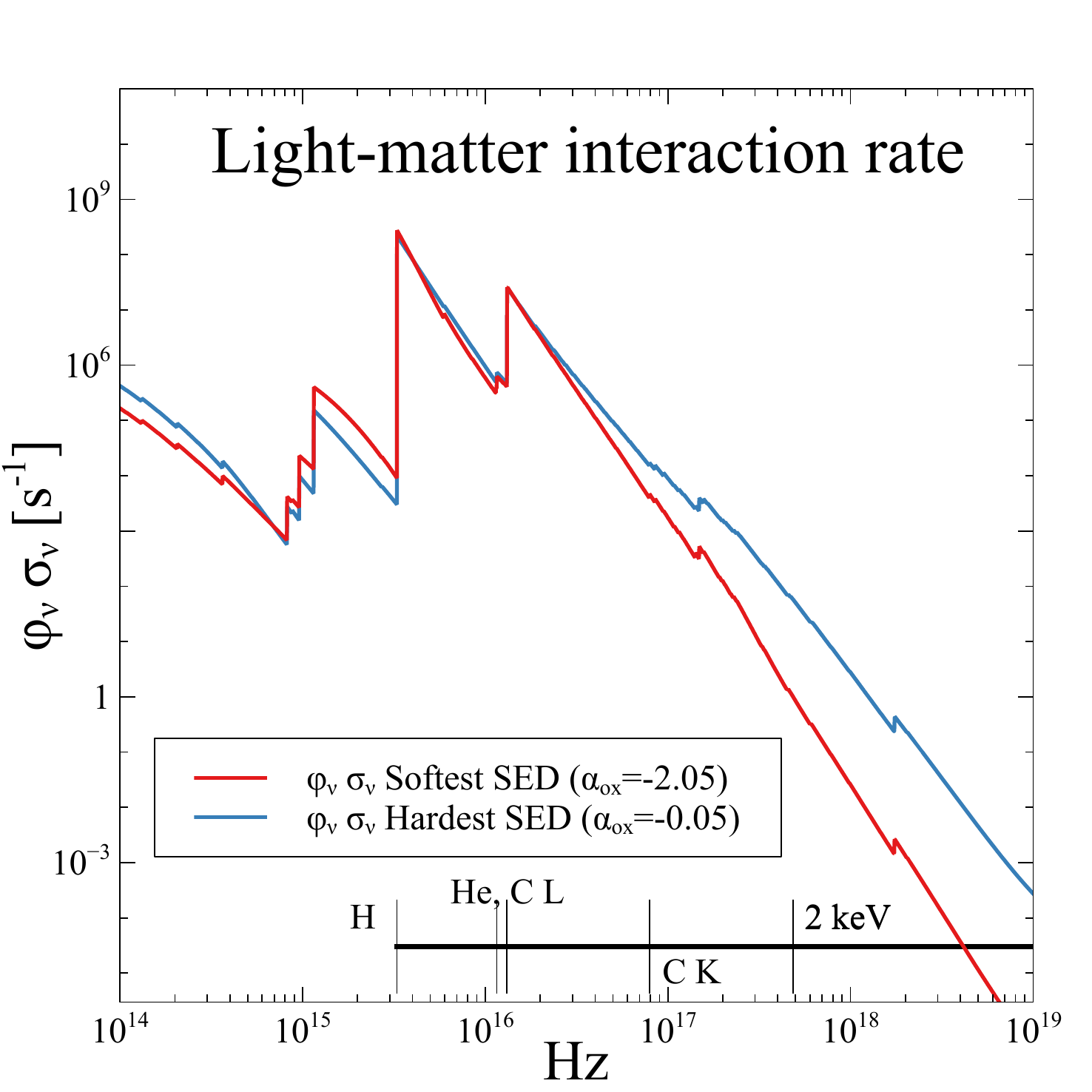}
    \caption{This shows the rate that photons interact with
    matter, the product of the flux of photons and the total
    gas opacity. 
    X-rays interact with matter with a rate about
    $\sim 7$ dex slower than the rate near the peak rate
    around $50 - 912$\AA.
    }
    \label{fig:PhiSigma}
\end{figure}

Figure \ref{fig:PhiSigma} shows the rate at which photons
interact with matter for our two reference
SEDs and a solar composition. 
Calculations are done with Cloudy version 22.01,
as last described by \citet{2017RMxAA..53..385F}.
Cloud parameters are typical of the \civ\ emitting
region of an AGN,
with a hydrogen density
of 10$^{10}$ cm$^{-3}$ and an ionization parameter
of $\log U = {-2}$.
The vertical axis is the total 
light-matter interaction
rate at a particular frequency and is the 
product of the photon flux and the total gas absorption
opacity, evaluated for the appropriate chemical
composition and degree of ionization.
The 50-912\,\AA\ ultraviolet region
%EUV - XUV - FUV regions,
is $\sim$7\,dex more interactive than
2\,keV X-rays.
As stated above, photoionization is photon counting,
and the relative paucity of X-ray photons cannot make up for
their greater energy.

%original data in Temple_Matthews_etal_22 / xray_sed_bis/PhiSigma.pdf
% veusz is SED.vsz sims are 027.in and 000.in 

High-energy photons would dominate the physics if
softer parts of the SED were extinguished
so only X-rays strike the gas.
Indeed, this is the 
`X-ray--dominated region' (XDR)
model of atomic and
molecular regions of clouds exposed to ionizing
radiation \citep{2022ARA&A..60..247W}.
It would be difficult to detect this XDR emission
since emission from the gas which absorbed the
softer radiation would be far stronger.
This is discussed in Section 4.1 of
\citet{2013RMxAA..49..137F}.
The ultraviolet luminosity of a realistic SED has more power 
than the relatively hard X-ray portion that drives an XDR.
The full SED striking a cloud produces successive 
H$^+$/H$^0$/H$_2$ layers, which are brighter 
than the deep X-ray heated regions.
Emission from regions powered by lower-energy light
would dominate over the XDR.
The penetrating X-rays, 
and galactic background cosmic rays,
are important for producing the small electron fraction that is present in deep regions of a molecular cloud.

\section{Additional observations}
\label{appendixd}

\begin{figure*}
	\includegraphics[width=\columnwidth]{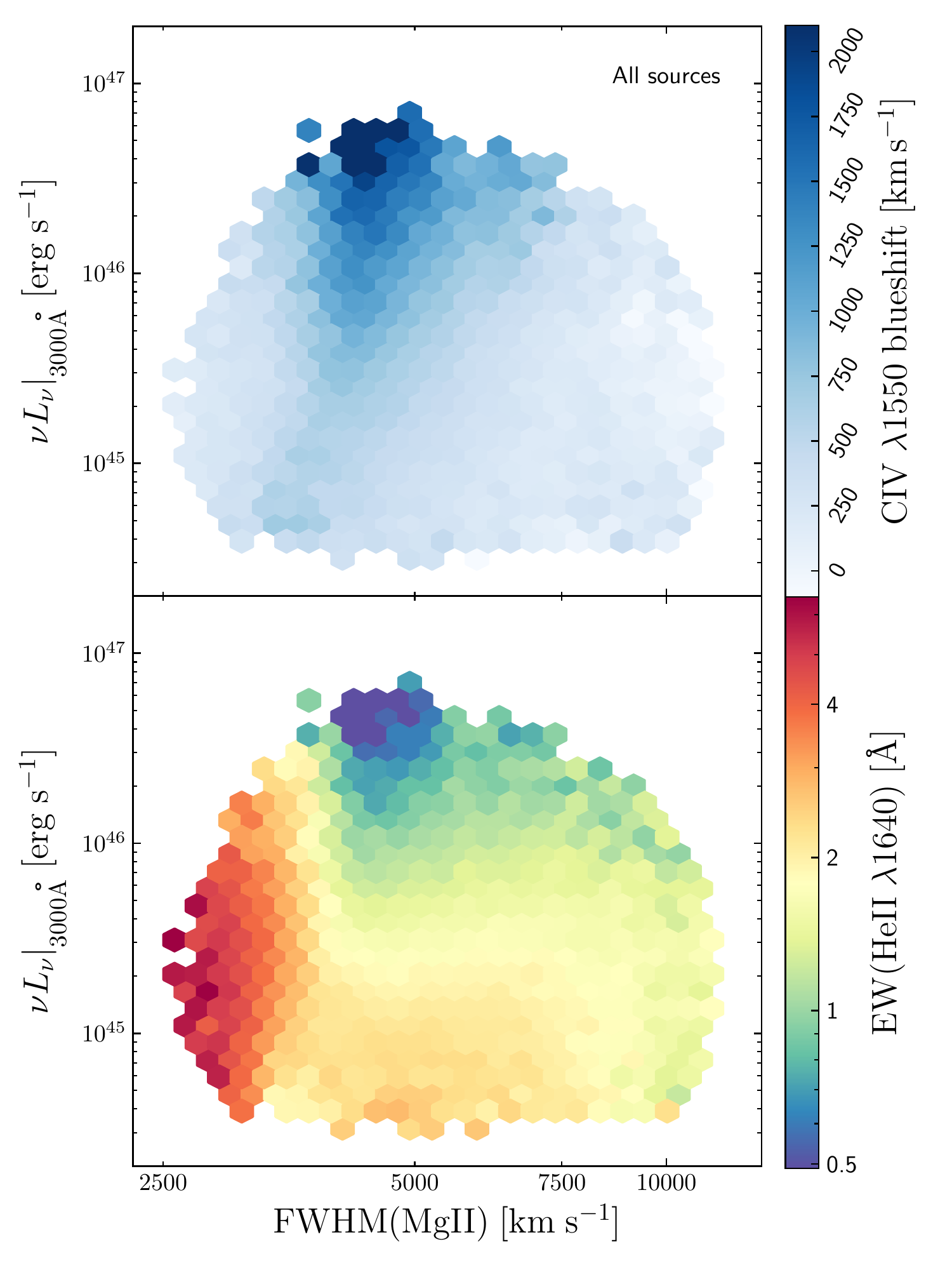}
	\includegraphics[width=\columnwidth]{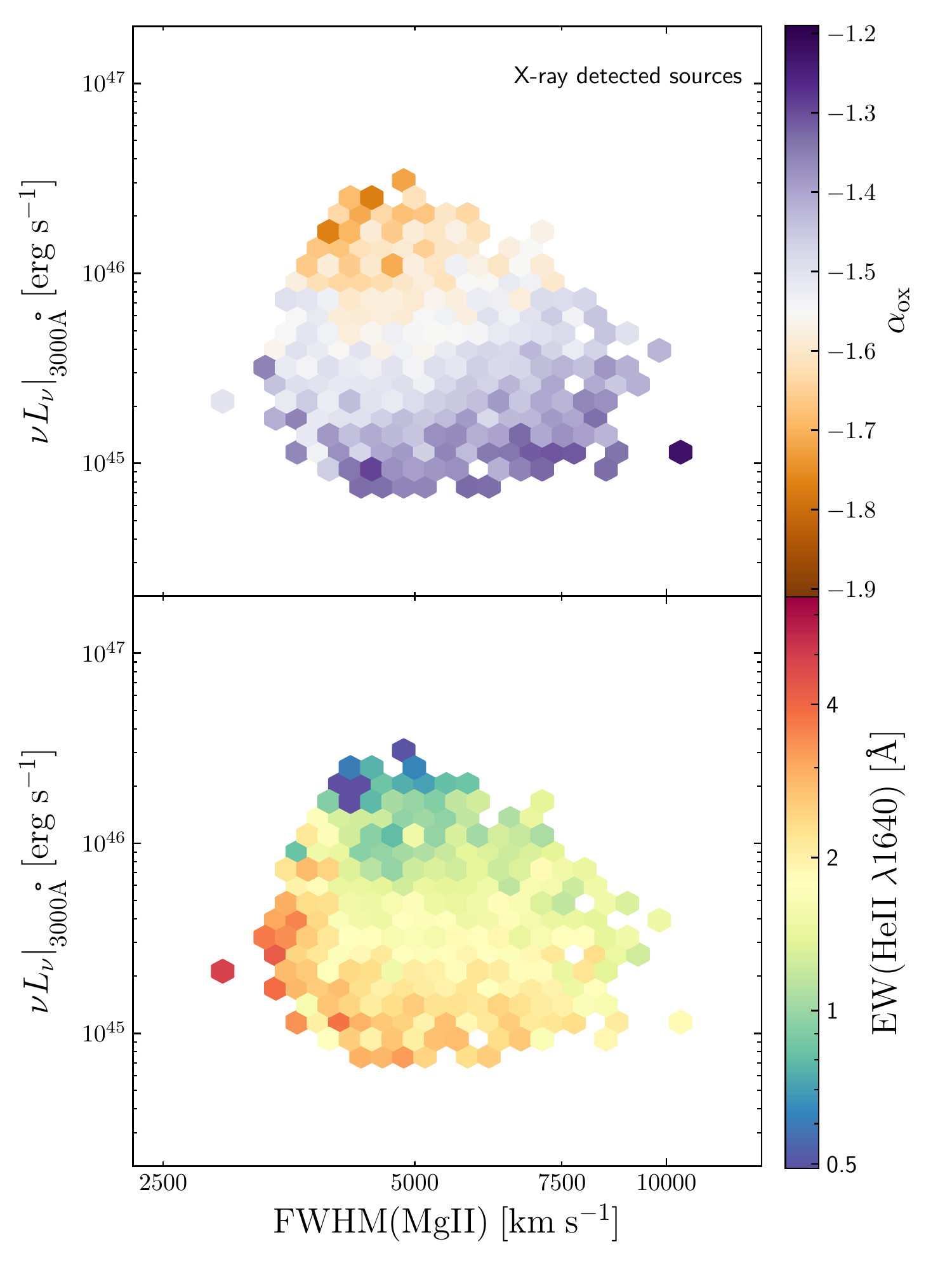}
    \caption{
    Observed quasar properties as a function of the FWHM of \mgii\,$\lambda2800$ and the 3000\,\AA\ continuum luminosity.
    \textsl{Left panel:}
    The median \ion{He}{II} EW (bottom) and \civ\ blueshift (top) in our full sample of 186\,303 objects. 
    \textsl{Right panel:} 
    The median \ion{He}{II} EW (bottom) and $\alpha_\textrm{ox}$ (top) in our  sub-sample of 5031 X-ray detected sources.
    The \ion{He}{II} behaviour is identical in both panels (modulo the sample size), suggesting that our X-ray detected sub-sample is not biased in terms of its ultraviolet emission properties.
    Moreover, clear differences are seen in the behaviour of \ion{He}{II} and $\alpha_\textrm{ox}$ within the X-ray subsample: the strongest \ion{He}{II} emission is seen only at low \mgii\ FWHM while the strongest 2\,keV X-ray emission is seen only at the lowest 3000\,\AA\ luminosities.
    }
    \label{fig:MgII}
\end{figure*}

\begin{figure*}
	\includegraphics[width=\columnwidth]{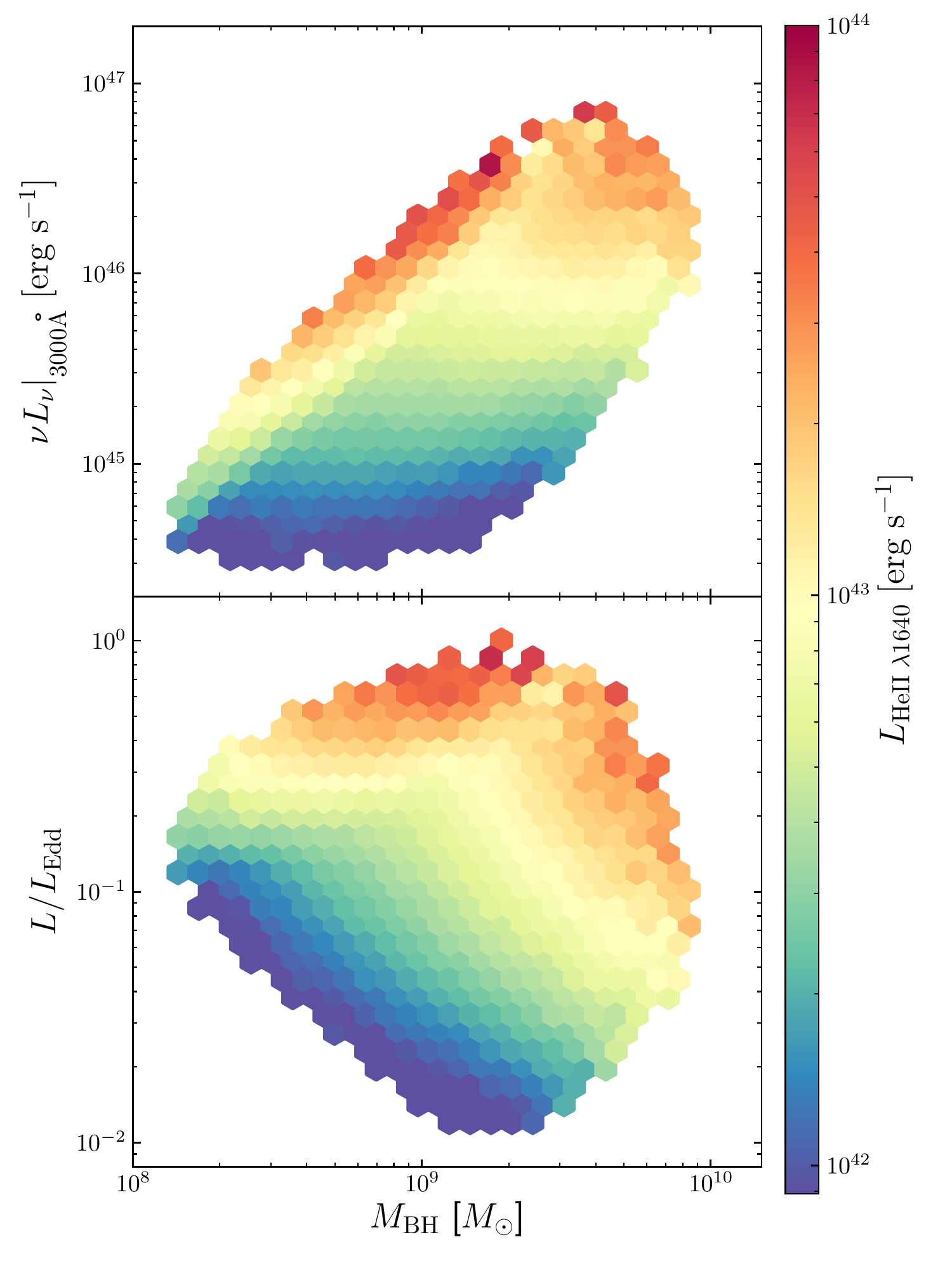}
	\includegraphics[width=\columnwidth]{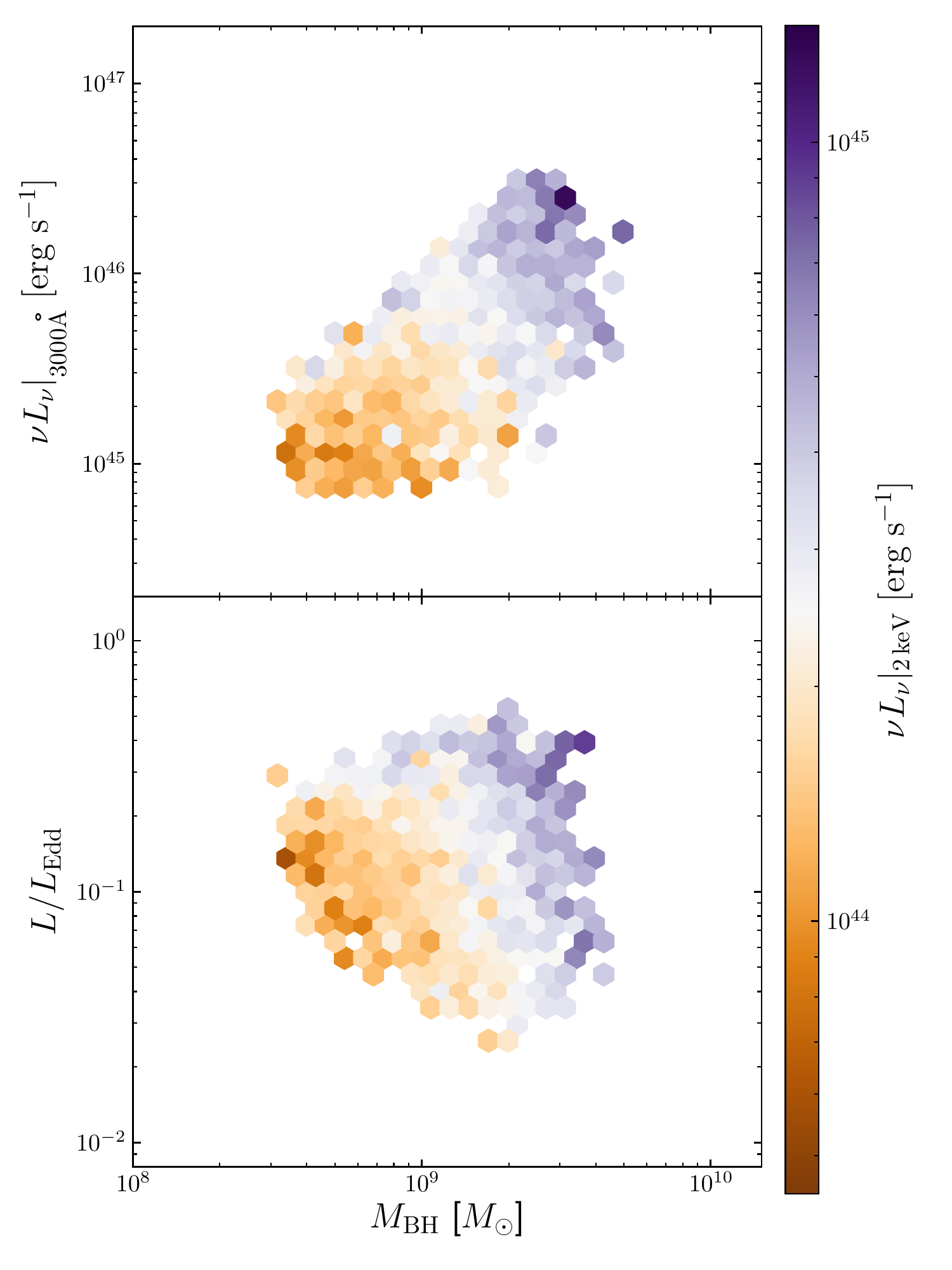}
    \caption{
    \ion{He}{II} line luminosity (left) and 2\,keV X-ray continuum luminosity (right) as a function of SMBH mass, luminosity and Eddington ratio.
    Contours of constant $L_\textrm{2\,keV}$ are largely aligned with lines of constant $M_\textrm{BH}$, consistent with the assumption in \qsosed\ that the hard X-ray emission is equal to $2$ per cent of the Eddington limit.
}
    \label{fig:Lums}
\end{figure*}

In this Appendix we present additional observational results.
First, in Fig.~\ref{fig:MgII} we present the \ion{He}{II} EW, the \civ\ blueshift and $\alpha_\textrm{ox}$ as a function of the FWHM of \mgii\,$\lambda2800$ and the 3000\,\AA\ continuum luminosity.
These two parameters are measured directly from the SDSS spectroscopy and photometry respectively. By contrast, the plots in the main text show observed properties as a function of 
\begin{equation}
M_\textrm{BH} \propto L_{3000}^{0.5}\textrm{FWHM}_\mgii^2
\end{equation}
and
\begin{equation}
L/L_\textrm{Edd} \propto L_{3000}/M_\textrm{BH} \propto L_{3000}^{0.5}\textrm{FWHM}_\mgii^{-2}.
\end{equation}
Given that both $M_\textrm{BH}$ and $L/L_\textrm{Edd}$ depend on the observed parameters $\textrm{FWHM}_\mgii$ and $L_{3000}$, this might lead to induced correlations in the $M_\textrm{BH}$--$L/L_\textrm{Edd}$ space. However, in practice our inferred $M_\textrm{BH}$--$L/L_\textrm{Edd}$ space is simply a rotation and reflection of the $\textrm{FWHM}_\mgii$--$L_{3000}$ space, where we still see clear trends.  
Furthermore, we see the same \ion{He}{II} behaviour  in the X-ray detected sub-sample as in our full sample, meaning that our X-ray detected objects are not obviously biased compared to our full sample.

Second, we show the 2\,keV X-ray continuum and \ion{He}{II}\,$\lambda1640$ line luminosities in Fig.~\ref{fig:Lums}. 
Assuming no changes in the BLR covering factor, and that the \ion{He}{II} continuum is optically thick, $L_\ion{He}{II}$ can be taken as a proxy for the continuum luminosity at 54\,eV.
These two observables show qualitatively different behaviour: contours of constant $L_\textrm{2\,keV}$ are largely aligned with lines of constant $M_\textrm{BH}$, which is consistent with the assumption in \qsosed\ that the hard X-ray power law component emits a constant fraction of the Eddington luminosity.
\ion{He}{II} behaves in a much more complex manner, with the gradient vector of increasing $L_\ion{He}{II}$ changing depending on the location in the $M_\textrm{BH}$--$L/L_\textrm{Edd}$ space.

\section{Bolometric corrections}
\label{appendixb}

\begin{figure*}
	\includegraphics[width=\textwidth]{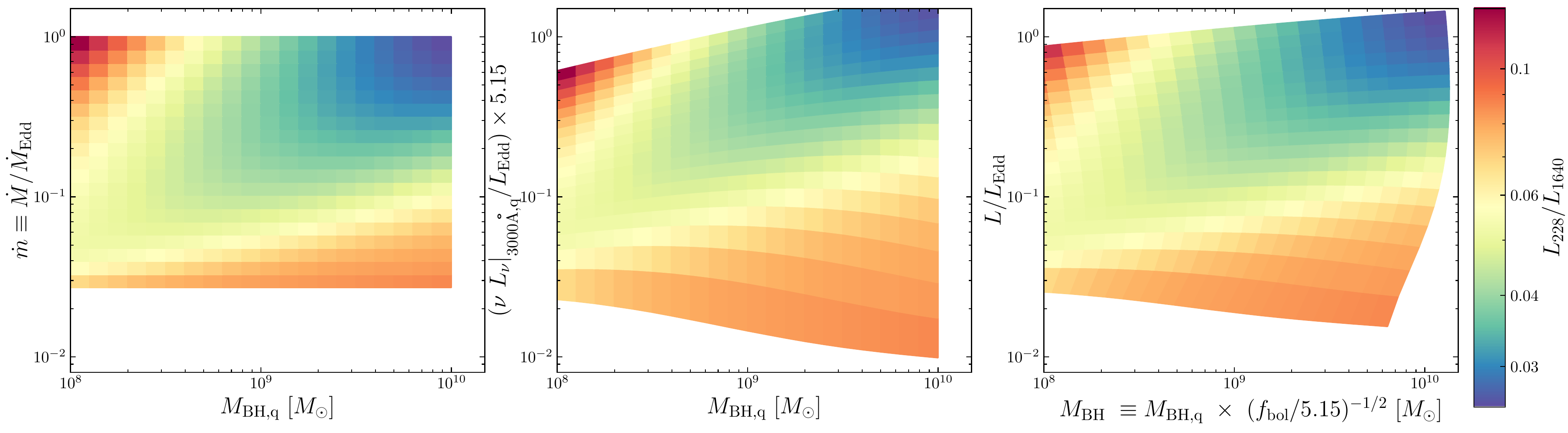}
    \caption{
    An illustration of how the \qsosed\ predictions change with differing treatments of the bolometric correction, focusing on the bottom-right panel of Fig.~\ref{fig:aox}. \textsl{Left panel:} the model outputs plotted as a function of the intrinsic, input values of $\dot{m}$ and $M_\textrm{BH, q}$. \textsl{Centre panel:} as in the left panel, but with the $y$-axis is replaced with $L_\textrm{bol}/L_\textrm{Edd}$ where $L_\textrm{bol}$ is calculated from $L_{3000}$ using a constant bolometric correction of $5.15$. \textsl{Right panel:} as in the centre panel, but with the $x$-axis scaled by $(f_\textrm{bol}/5.15)^{-1/2}$ and the $y$-axis scaled by $(f_\textrm{bol}/5.15)^{1/2}$ to capture the impact of the bolometric correction on SMBH mass estimates in the observed data. See the main Appendix text for details.  
    }
    \label{fig:f_bol}
\end{figure*}

We have applied a fixed bolometric correction of $5.15$ to estimate $L_\textrm{bol}$ from $\nu L_\nu|_{\textrm{3000\AA}}$. In reality, the bolometric correction will vary as a function of $M_\textrm{BH}$ and $L/L_\textrm{Edd}$. We discussed the variation of the bolometric correction from the \qsosed\ model grid in Section~\ref{sec:bolometric_corr}, showing a range in $f_\textrm{bol}$ by a factor of $\approx 2-3$, where $f_\textrm{bol}\equiv \nu L_\nu|_\textrm{3000\AA} / L_\textrm{bol}$ is calculated from each individual \qsosed\ model. Although a true `Apples versus Apples' comparison is only really possible with full knowledge of the intrinsic SED, in comparing our \qsosed\ models with data we tried to match scalings and biases in the data introduced by the fixed bolometric correction by applying appropriate transformations to the \qsosed\ outputs. Our single-epoch virial SMBH mass estimates make use of the observed \mgii\ line width, but also require an estimate of the line formation radius for which we follow the usual method and assume that the BLR radius scales as $R_\textrm{BLR} \propto L^{1/2}$. The $L$ in this expression should really be some appropriate ionizing luminosity, but $L_\textrm{bol}$ is normally used and we follow this convention. As a result, the bolometric correction enters into the SMBH mass estimate and implies a bias in the SMBH mass estimates with respect to the true SMBH mass by factor of $(f_\textrm{bol}/5.15)^{-1/2}$. As a result, when plotting $M_\textrm{BH}$ along the $x$-axis of Figs.~\ref{fig:HeII} and \ref{fig:aox}, we apply the scaling
\begin{equation}
M_\textrm{BH} = (f_\textrm{bol}/5.15)^{-1/2} M_\textrm{BH, q},
\end{equation}
where $M_\textrm{BH, q}$ denotes the input \qsosed\ grid value (the `true' SMBH mass). For $L/L_\textrm{Edd}$, correction factors appear in both the numerator and denominator. $L_\textrm{Edd} \propto M_\textrm{BH}$, introducing a bias factor $(f_\textrm{bol}/5.15)^{-1/2}$ into the Eddington ratio estimate, while the numerator is $L_\textrm{bol}$ and so contains a straightforward bias factor of $f_\textrm{bol}/5.15$. As a result, the relationship between the $L/L_\textrm{Edd}$ plotted in Figs.~\ref{fig:HeII} and \ref{fig:aox}, and the dimensionless, Eddington-scaled accretion rate used as input to \qsosed\ is given by
\begin{equation}
L/L_\textrm{Edd} = \dot{m} \times (f_\textrm{bol}/5.15) \times (f_\textrm{bol}/5.15)^{-1/2}. 
\end{equation}
The effect of introducing these scaling factors as transformations from the initial \qsosed\ grid is shown in Fig.~\ref{fig:f_bol}, to show how the right-hand panels of Figs.~\ref{fig:aox} and \ref{fig:HeII} would change if we had made a different presentation choice. The scale factors twist and distort the simulation grid slightly from the original uniform parameter space, but, overall, the effects are fairly modest because only square-root terms distinguish the rightmost panel from the original input grid.

\section{Black hole spin}
\label{appendixc}

\begin{figure*}
        \includegraphics[width=0.33\textwidth]{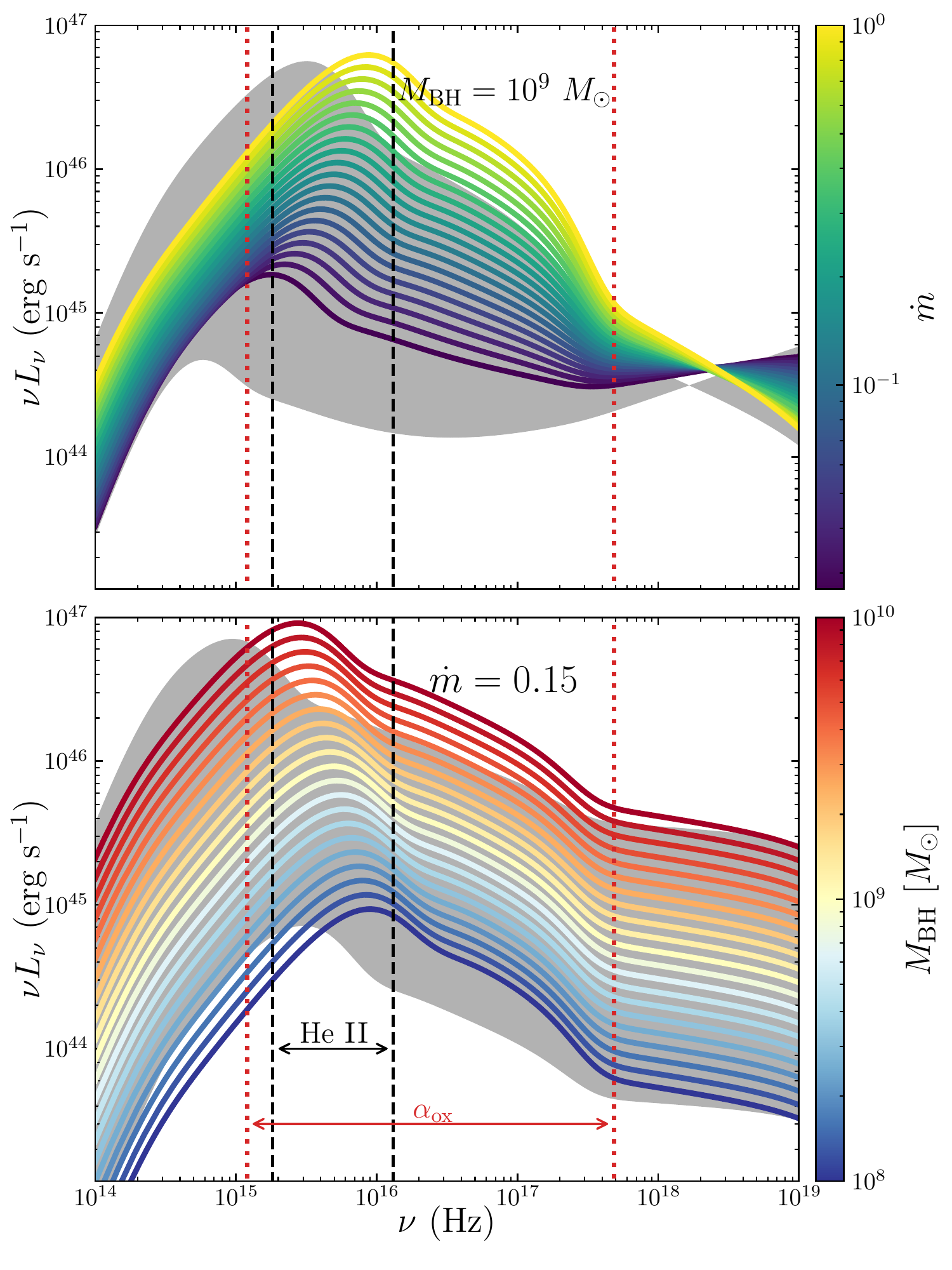}
	\includegraphics[width=0.33\textwidth]{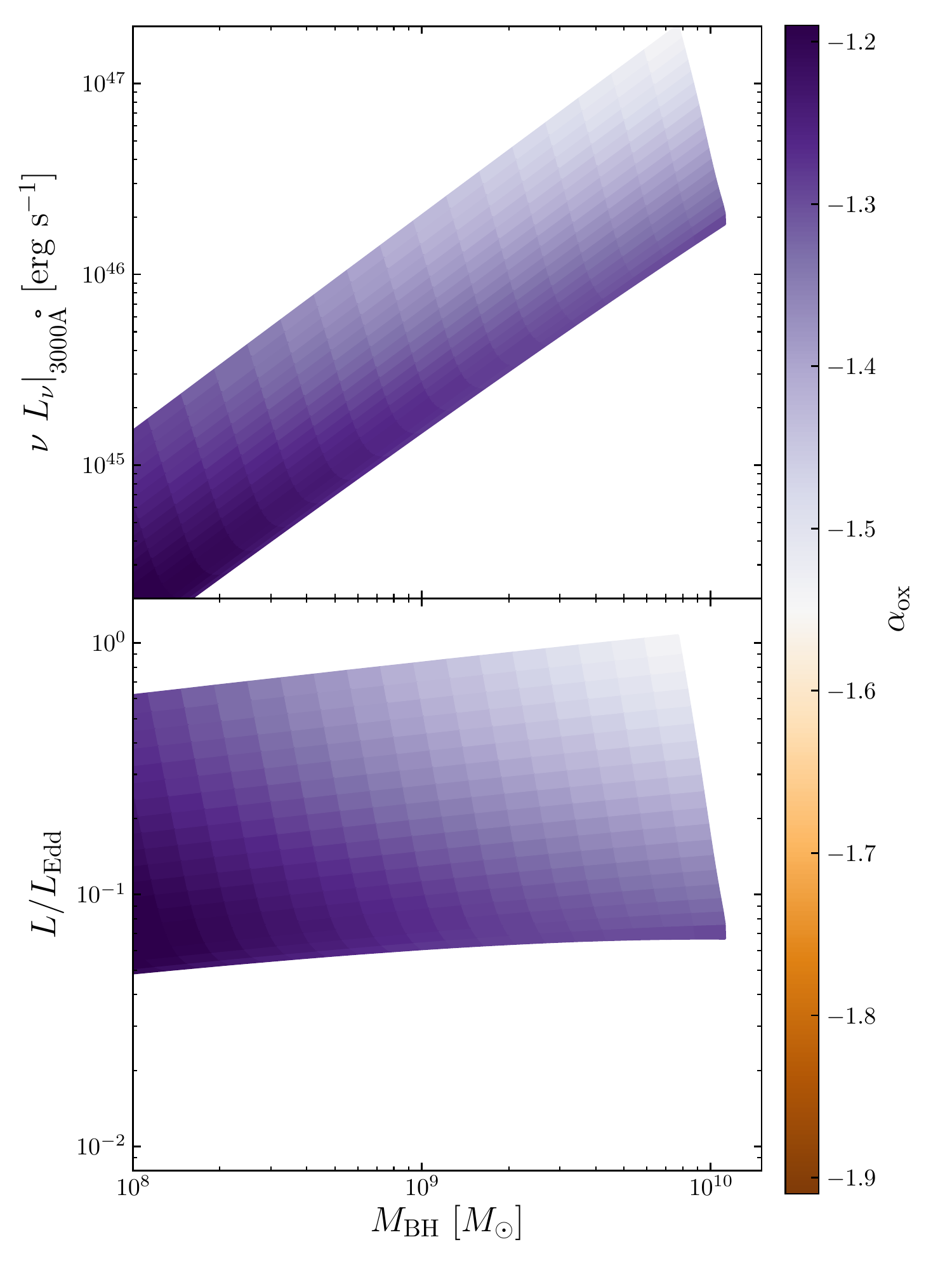}
	\includegraphics[width=0.33\textwidth]{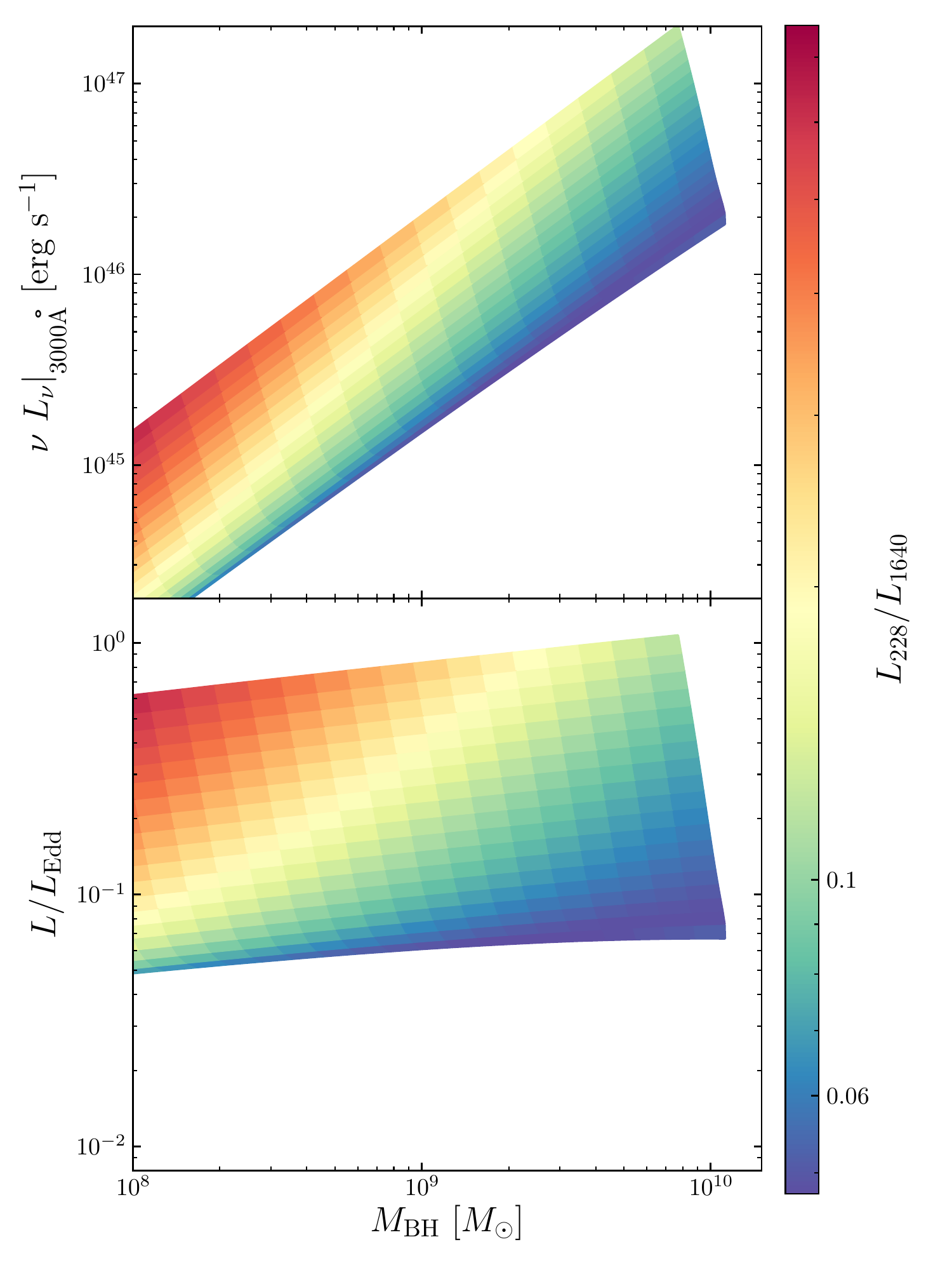}
    \caption{
    \textsl{Left panel:} As Fig.~\ref{fig:qso_seds}, but for the maximally spinning case, $a_*=0.998$. The grey shaded area shows the range of the SEDs shown in Fig.~\ref{fig:qso_seds} for the non-spinning case.
    \textsl{Centre and right panels:} \qsosed\ predictions for $\alpha_\textrm{ox}$ and $L_{228}/L_{1640}$ for a maximally spinning SMBH with $a_*=0.998$ (cf. Figs.~\ref{fig:aox} and \ref{fig:HeII}). 
    }
    \label{fig:bh_spin}
\end{figure*}

In Fig.~\ref{fig:bh_spin} we show how the predictions of $\alpha_\textrm{ox}$ and our \ion{He}{II}\,$\lambda$1640 EW proxy ($L_{228}/L_{1640}$) from \qsosed\ change if we instead consider a maximally spinning SMBH. While the qualitative trends in the $L_{228}/L_{1640}$ are broadly in line with the low spin case, the $a_*=1$ models fail to reproduce the observed low values of $\alpha_\textrm{ox}$ at high Eddington fractions and SMBH masses (see discussion in Section~\ref{sec:spin_inc}). The reason for this can be understood from the left-hand panel of Fig.~\ref{fig:bh_spin}, where we show the \qsosed\ broadband spectrum (the analogue to Fig.~\ref{fig:qso_seds}) for the maximally spinning case. Inspection of the high $\dot{m}$ models in the top-panel reveals that the $\alpha_\textrm{ox}$ behaviour is driven by a combination of stronger X-rays and the movement of the peak of the thermal component. At high spin, the thermal peak moves blueward to higher energies, such that the lower frequency pivot point falls further from the peak and has lower flux compared to the low spin model. In \qsosed, this behaviour comes about in a slightly convoluted way, but is driven by the decrease of the radius $R_\textrm{warm}$ (and corresponding temperature increase). This decrease happens because $R_\textrm{warm} = 2 R_\textrm{hot}$, and $R_\textrm{hot}$ must move inwards as spin increases, because $R_\textrm{ISCO}$ moves closer to the SMBH  so $R_\textrm{hot}$ must also decrease from eq.~2 of \cite{2018MNRAS.480.1247K} to maintain the model assumption that the dissipated power is $2$ per cent of the Eddington luminosity. One could clearly construct other models in which the critical radii change in different ways when the spin is changed, which is partly why we caution against over-interpreting the fact that maximal spins appear difficult to reconcile with the data.

%%%%%%%%%%%%%%%%%%%%%%%%%%%%%%%%%%%%%%%%%%%%%%%%%%

% Don't change these lines
\bsp	% typesetting comment
\label{lastpage}
\end{document}